\documentclass[preprint2]{aastex}
\shorttitle{Polarimetry of NGC 2264 Region}
\shortauthors{Kwon et al.}

\newcommand{\mcoc}[1]{\multicolumn{1}{c}{{#1}}}

\usepackage{times}
\slugcomment{To appear in the Astrophysical Journal}

\begin{document}

\fontsize{10}{10.6}\selectfont
\title{Complex Scattered Radiation Fields and Multiple Magnetic Fields \\ 
       in the Protostellar Cluster in NGC 2264}

\author{\sc Jungmi Kwon\altaffilmark{1,2}, 
            Motohide Tamura \altaffilmark{1,2}, Ryo Kandori \altaffilmark{1}, 
            Nobuhiko Kusakabe\altaffilmark{1}, Jun Hashimoto\altaffilmark{1}, 
            Yasushi Nakajima\altaffilmark{1}, Fumitaka Nakamura\altaffilmark{1}, 
            Takahiro Nagayama\altaffilmark{3}, Tetsuya Nagata\altaffilmark{4}, 
            James H. Hough\altaffilmark{5}, Michael W. Werner\altaffilmark{6}, 
            and Paula S. Teixeira\altaffilmark{7}}
\affil{$^1$ National Astronomical Observatory of Japan, 2-21-1 Osawa, Mitaka, Tokyo 181-8588, Japan; jungmi.kwon@nao.ac.jp}
\affil{$^2$ Department of Astronomical Science, Graduate University for Advanced Studies (Sokendai), 2-21-1 Osawa, Mitaka, Tokyo 181-8588, Japan }
\affil{$^3$ Department of Astrophysics, Nagoya University, Nagoya 464-8602}
\affil{$^4$ Department of Astronomy, Kyoto University, Kyoto 606-8502}
\affil{$^5$ Centre for Astrophysics Research, University of Hertfordshire, Hatfield, Herts AL10 9AB, UK}
\affil{$^6$ Jet Propulsion Laboratory, California Institute of Technology, Pasadena, CA 91109, USA}
\affil{$^7$ European Southern Observatory, Karl-Schwarzschild-Stra\ss e 2, D-85748 Garching bei M\"{u}nchen, Germany}

\begin{abstract}

\fontsize{10}{10.6}\selectfont

Near-infrared (IR) imaging polarimetry in the $J$, $H$, and $K_s$ bands
has been carried out for the protostellar cluster region around NGC 2264 IRS 2 
in the Monoceros OB1 molecular cloud.
Various infrared reflection nebulae clusters (IRNCs)
associated with NGC 2264 IRS 2 and IRAS 12 S1 core were detected
as well as local infrared reflection nebulae (IRNe). 
The illuminating sources of the IRNe were identified with known or new near- and mid-IR sources. 
In addition, 314 point-like sources were detected in all three bands 
and their aperture polarimetry was studied.
Using a color--color diagram, 
reddened field stars and diskless pre-main sequence stars were selected
to trace the magnetic field (MF) structure of the molecular cloud.
The mean polarization position angle of the point-like sources 
is 81$\arcdeg$ $\pm$ 29$\arcdeg$ in the cluster core,
and 58$\arcdeg$ $\pm$ 24$\arcdeg$ in the perimeter of the cluster core,
which is interpreted as the projected direction on the sky of the MF 
in the observed region of the cloud.
The Chandrasekhar--Fermi method
gives a rough estimate of the MF strength
to be about 100 $\mu$G.
A comparison with recent numerical simulations of the cluster formation 
implies that the cloud dynamics is controlled by the relatively strong MF.
The local MF direction is well associated with that of CO outflow for IRAS 12 S1 
and consistent with that inferred from submillimeter polarimetry.
In contrast, the local MF direction runs roughly perpendicular to the Galactic MF direction.
\end{abstract}


\section{INTRODUCTION}

NGC 2264, including the Cone Nebula, Fox Fur Nebula and Christmas Tree Cluster 
in the Monoceros OB1 molecular cloud, 
is a site of active star formation (Walker 1954, 1956)
and is a cornerstone
for the study of the formation and time evolution of young stellar objects (YSOs).
The median age of NGC 2264 is 1--3 Myr 
with an age dispersion of $\sim$5 Myr (Dahm \& Simon 2005). 
It is located at a distance of $\sim$760 pc (Sung et al. 1997, but see also Baxter et al. 2009, Sung et al. 2010) 
and lies 37 pc above the galactic plane (P\'{e}rez 1991)
in front of a dark molecular cloud complex.
The mass of the cloud complex is estimated 
to be at least $\sim$3.7 $\times$ 10$^4$ M$_{\sun}$ (Dahm 2008).
NGC 2264 has an intermediate stellar density 
and its stellar population, which is known down to very low masses,
is dominated by the O7V multiple star S Monocerotis 
(Schwartz et al. 1985). 

The NGC 2264 region has been exceptionally well studied at various wavelengths 
from centimeters to X-rays.
Previous studies have revealed that 
active star formation is ongoing in this region,
as evidenced by the presence of numerous embedded clusters 
of protostars, molecular outflows, and Herbig--Haro objects
(e.g., Adams et al. 1979; Margulis et al. 1989; 
Walsh et al. 1992; Hodapp 1994; Reipurth et al. 2004; Young et al. 2006, hereafter YTL). 
There are two prominent sites of star formation activity 
centered near NGC 2264, identified by IR and millimeter observations,
IRAS 06384+0932 (hereafter IRS 1) and IRAS 06382+0939 (hereafter IRS 2). 
NGC 2264 IRS 1, known as Allen's source (Allen 1972), 
is a deeply embedded early-type (B2-B5) star. 
NGC 2264 IRS 2, discovered by Sargent et al. (1984), was identified as a Class I source.
NGC 2264 IRS 2 was designated as IRAS 12
(06$^h$ 41$^m$ 02$\fs$7 +09$\arcdeg$ 36$\arcmin$ 10$\arcsec$) 
by Margulis et al. (1989),
and Castelaz \& Grasdalen (1988) showed that
NGC 2264 IRS 2 is a binary source (RNO-E and RNO-W) separated by 2$\farcs$8.

Kr\"{u}gel et al. (1987) presented an extended ammonia map 
of the central region of the molecular cloud 
associated with NGC 2264 including both IRS 1 and IRS 2.
Williams \& Garland (2002) also observed the dust and gas 
near the two young stellar clusters IRS 1 and IRS 2 in NGC 2264
and presented an 870 $\mu$m continuum emission map.
The elongated shape of IRS 1 shows signs of substructure, 
while IRS 2 is more fragmented, indicating a more evolved cluster of protostars.
IRAS 12 S1 located south-west of NGC 2264 IRS 2 was found by Cohen et al. (1985),
and it is located precisely at the peak of the IRAS 12 S1 
in the submillimeter maps of Wolf-Chase et al. (2003).
It is also associated with the core C of Williams \& Garland (2002), 
the NGC 2264 D-MM1 of Peretto et al. (2006),
and the dense microcluster of Class 0 protostars of Teixeira et al. (2007).
Wolf-Chase et al. (2003) derived an envelope mass of 17.6 M$_{\sun}$ within a 29$\arcsec$ diameter,
and Peretto et al. (2006) 
estimated a mass infall rate of \.{M}$_{D-MM1}$ $\sim$1.1 $\times$ 10$^{-4}$ M$_{\sun}$yr$^{-1}$ 
from millimeter observations, toward the rotating IRAS 12 S1. 
However, 
YTL and Teixeira et al. (2007) showed that
IRAS 12 S1 is in fact a multiple source, composed of at least 7 protostars.

Teixeira et al. (2006) identified
bright 24 $\mu$m protostars that are radially distributed around NGC 2264 IRS 2,
and embedded within dusty filaments.
The cluster of protostars was therefore named the Spokes cluster.
The projected separation of the 24 $\mu$m protostars is 20 $\pm$ 5$\arcsec$,
similar to the Jeans length of the cloud, 27\arcsec,
implying the filaments had undergone thermal fragmentation.
In addition, the evolutionary stages of protostars in the NGC 2264 IRS 2 region
were studied through detailed spectral energy distributions (SEDs)
(Forbrich et al. 2010, hereafter FTR), 
and the outflow activity was derived in the vicinity of IRAS 12 S1.
Bourke et al. (2001), from OH Zeeman observations 
using the NRAO Green Bank 43-m telescope (FWHM beam size $\sim$18$\arcmin$),
estimated the magnetic field strength for NGC 2264 to be 21 $\pm$ 16 $\mu$G.

The outflows and the Herbig--Haro objects are considered as indicators 
of recent star formation activity.
The characteristics of outflows and jets associated with YSOs
are related to the magnetic field
and the rotation of the protostar and circumstellar disk.
The outflows and jets commonly tend to be aligned
with each other and with the cloud-scale magnetic field,
as determined from polarization observations 
(e.g., Cohen et al. 1984; Strom et al. 1986; Vrba et al. 1986, 1988; Tamura \& Sato 1989; Jones \& Amini 2003). 
Herbig (1974) performed early surveys of the NGC 2264 region
and identified several candidate Herbig--Haro objects, 
as well as a survey of emission line stars (Herbig 1954).

Multi-color near-IR polarimetry is useful 
for understanding magnetic fields and the properties of dust grains 
that cause scattering and absorption in various environments
(e.g., Tamura et al. 2007).
Polarimetric studies of the NGC 2264 region have been reported previously by several authors
(e.g., Breger \& Hardorp 1973; Kobayashi et al. 1978; Dyck \& Lonsdale 1979; Heckert \& Zeilik 1981, 1984; Schreyer et al. 2003; Dotson et al. 2010)
but almost all of these studies only covered the NGC 2264 IRS 1 region.
The polarization position angle of NGC 2264 IRS 1 was found to be  $\sim$106$\arcdeg$--117$\arcdeg$
in the near-IR (Breger \& Hardorp 1973; Kobayashi et al. 1978; Heckert and Zeilik 1981).
Dotson et al. (2010) recently reported 350 $\mu$m polarimetry 
in the direction of IRS 1 and IRAS 12 S1 near IRS 2 in NGC 2264.

In this paper, we first present wide-field near-IR imaging polarimetry of the IRS 2 region, 
as part of our ongoing survey project of $JHK_s$ polarimetry 
for star forming regions.
In Section 2, we describe the observations.
In Section 3, we present the results of the imaging and aperture polarimetry.
In Section 4, we discuss the illuminating sources of infrared reflection nebulae
and the magnetic field structure 
related to the star forming activity in the NGC 2264 IRS 2 region.
A summary is given in Section 5.

\section{OBSERVATIONS}

\begin{figure*}
\epsscale{1.9}
\plotone{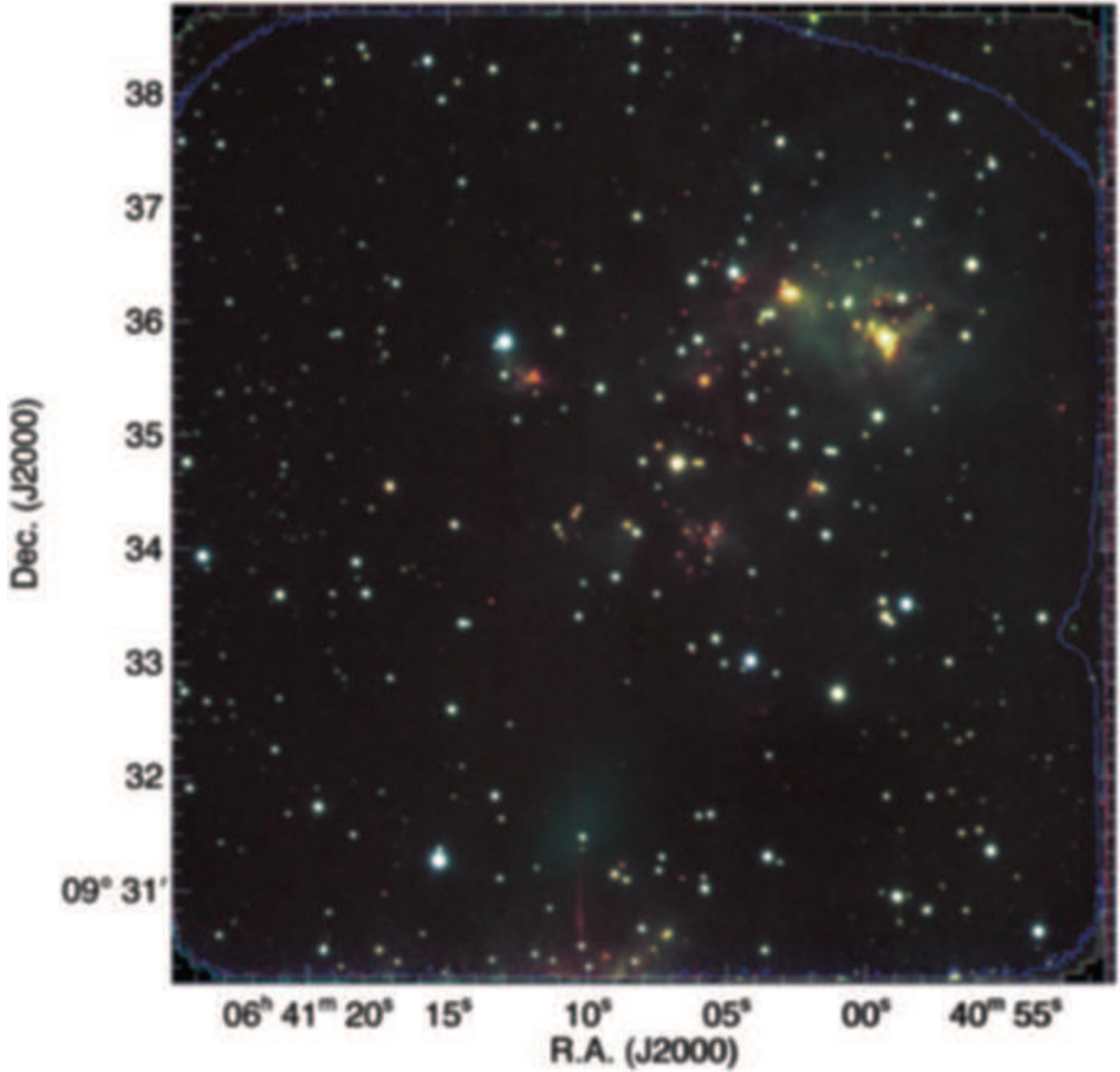}
\vspace{-0.5\baselineskip}
\vspace{-0.5\baselineskip}
\caption{
Color composite Stokes $I$ image of the IRS 2 region
in the $J$ ({\it blue}), $H$ ({\it green}), and $K_s$ ({\it red}) bands
from the IRSF/SIRPOL observations.
The thin lines of red, green, and blue around the perimeter are boundaries of Stokes $I$ images 
associated with each band.
Note that there are bad pixel clusters around the upper-left and upper-right corners 
and the middle of the right boundary.
See Figure 2 for identification of selected sources.}
\end{figure*}

\begin{figure*}
\epsscale{1.9}
\plotone{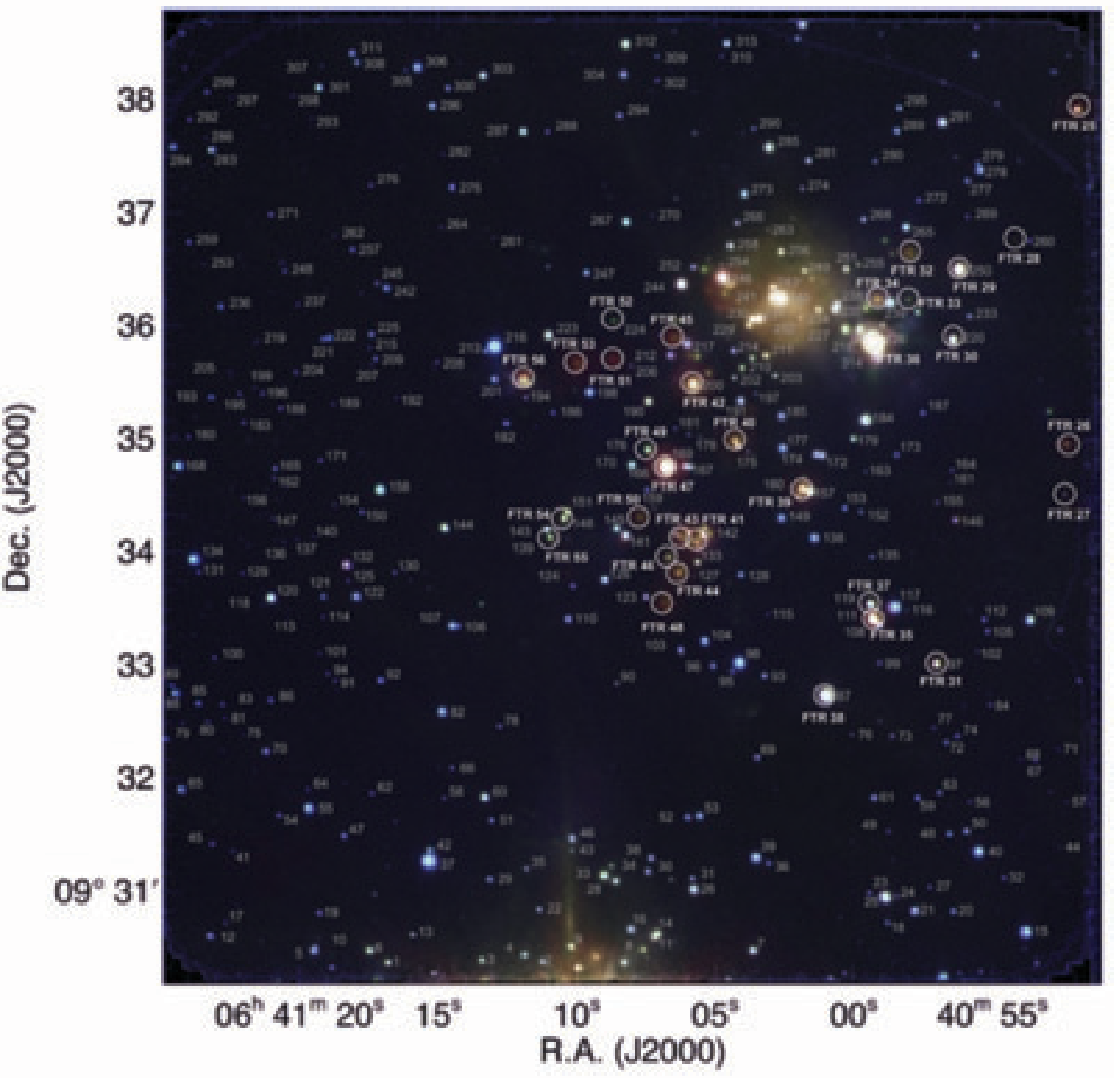}
\vspace{-0.5\baselineskip}
\vspace{-0.5\baselineskip}
\caption{
Finding chart of the IRS 2 field on the color composite image 
in the IRSF/SIRPOL $JHK_s$ ({\it blue}),
Spitzer/IRAC 5.8 $\mu$m ({\it green}),
and Spitzer/MIPS 24 $\mu$m ({\it red}) bands.
Detected point-like sources are labeled (Table 1).
24 $\mu$m sources from Forbrich et al. (2010) are denoted by both white circles and labels.}
\end{figure*}

The observations in the direction of the IRAS 12 S1 region were carried out
using the SIRPOL imaging polarimeter
on the Infrared Survey Facility (IRSF) 1.4-m telescope
at the South African Astronomical Observatory.
SIRPOL consists of a single-beam polarimeter
(an achromatic half-wave plate rotator unit and a polarizer)
and an imaging camera (Nagayama et al. 2003).
The camera, SIRIUS, has three 1024 $\times$ 1024 HgCdTe infrared detectors.
IRSF/SIRPOL enables deep and wide-field
(7\farcm7 $\times$ 7\farcm7 with a scale of 0\farcs45 pixel$^{-1}$)
imaging polarimetry in the $J$, $H$, and $K_s$ bands simultaneously
(Kandori et al. 2006).

The observations were made on the night of 2007 February 19.
We performed 10-s exposures at four wave-plate angles
(in the sequence 0\arcdeg, 45\arcdeg, 22$\fdg$5, and 67$\fdg$5)
at 10 dithered positions for each set.
The same observation sets were repeated 10 times
toward the target object and for the sky background
to increase the signal-to-noise ratio.
The total integration time was 1000 s per wave plate angle.
The typical seeing size during the observations
was $\sim$1\farcs3 (2.9 pixels) in the $J$ band.
SIRPOL has been routinely used since 2006
and the instrumental polarization is negligible (Kandori et al. 2006).
The polarization efficiencies are 95.5 \%, 96.3 \%, and 98.5 \% 
in the $J$, $H$, and $K_s$ bands, respectively.

The data were processed using IRAF in the standard manner,
which included dark-field subtraction, flat-field correction,
median sky subtraction, and frame registration.
The remaining artificial stripe pattern was then removed using IDL.
The pixel coordinates of point-like sources found on the reduced images 
were matched with the celestial coordinates of their counterparts
in the Two Micron All Sky Survey (2MASS) point source catalog.
The IRAF IMCOORDS package was applied to the matched list
to obtain plate transform parameters.
The rms uncertainty in the coordinate transformation was $\sim$0\farcs1.

Figure 1 shows the $J$-$H$-$K_s$ color composite intensity image
of the 8$\farcm$0 $\times$ 8$\farcm$0 region around NGC 2264 IRS 2 
(hereafter the IRS 2 field) 
including IRAS 12 S1 core.
The field is larger than the field-of-view of the camera because of dithering.

Mid-IR data obtained from Spitzer IRAC and MIPS were used for data analysis as well as our near-IR data. 
The IRAC 5.8 $\mu$m data were acquired in two epochs 7 months apart (2004 March 6 and October 8), 
with two dithers at each epoch to allow easy removal of asteroids and other transients, 
as part of Spitzer Guaranteed Time Observation program 37. 
Basic data reduction and calibration were done with the Spitzer Science Center (SSC) pipeline, version S14.0. 
The MIPS 24 $\mu$m data were obtained on 2004 March 16 using the scan-map mode, 
as part of Spitzer Guaranteed Time Observation program 58,
and the data were reduced using the MIPS Data Analysis Tool, version S9.5. 
Since the spatial resolution of the MIPS 24 $\mu$m data is low, 
we paid considerable attention to finding the appropriate IRAC 5.8 $\mu$m and SIRPOL counterparts.

Figure 2 shows the $J$-$H$-$K_s$--5.8 $\mu$m--24 $\mu$m
color composite intensity image of the IRS 2 region.
Since protostars have significant circumstellar disks and envelopes,
the 24 $\mu$m band may be particularly useful
for identifying the Class I and Class 0 objects.

In this paper, 
we designate individual infrared reflection nebula(e) as IRN(e)
and clusters of those infrared reflection nebulae as IRNC(s).
We also designate the IRN associated with an IR source X as IRN (X)
such as IRN (AR 6), IRN (IRS 2), and IRN (D-MM15), rather than numbering each.

\section{RESULTS}

\subsection{Polarimetry of extended sources}

Polarimetry of extended sources was carried out
on the combined intensity images for each exposure cycle
(a set of exposures at four wave plate angles at the same dithered position).
The Stokes parameters $I$, $Q$, and $U$ were calculated by
\begin{equation}
   I = {1 \over 2} (I_{0} + I_{22.5} + I_{45} + I_{67.5}),
\end{equation}
\begin{equation}
   Q = I_{0} - I_{45},
\end{equation}
and
\begin{equation}
   U = I_{22.5} - I_{67.5},
\end{equation}
where $I_a$ is the intensity with the half wave plate oriented at $a\arcdeg$.
The polarization degree $P$ and the polarization position angle $\theta$ were calculated by
\begin{equation}
   P = {{\sqrt{Q^2 + U^2}} \over {I}}
\end{equation}
and
\begin{equation}
   \theta = {1\over2} \arctan {U \over Q}.
\end{equation}
The degree of polarization $P$ was corrected using the polarization efficiencies of SIRPOL.

Figure 3 shows the $H$ polarization vector map of the whole IRS 2 region
superposed on the intensity $I$ image combined with SIRPOL$+$Spitzer image data.
This figure shows prominent and extended polarization nebulosities
over the IRS 2 region for the first time.
The polarimetry of extended sources can reveal 
the locations of illuminating sources within nebulae
because the observed polarization is perpendicular to ray from illuminating source 
to scatterer. 

\begin{figure*}
\epsscale{1.9}
\plotone{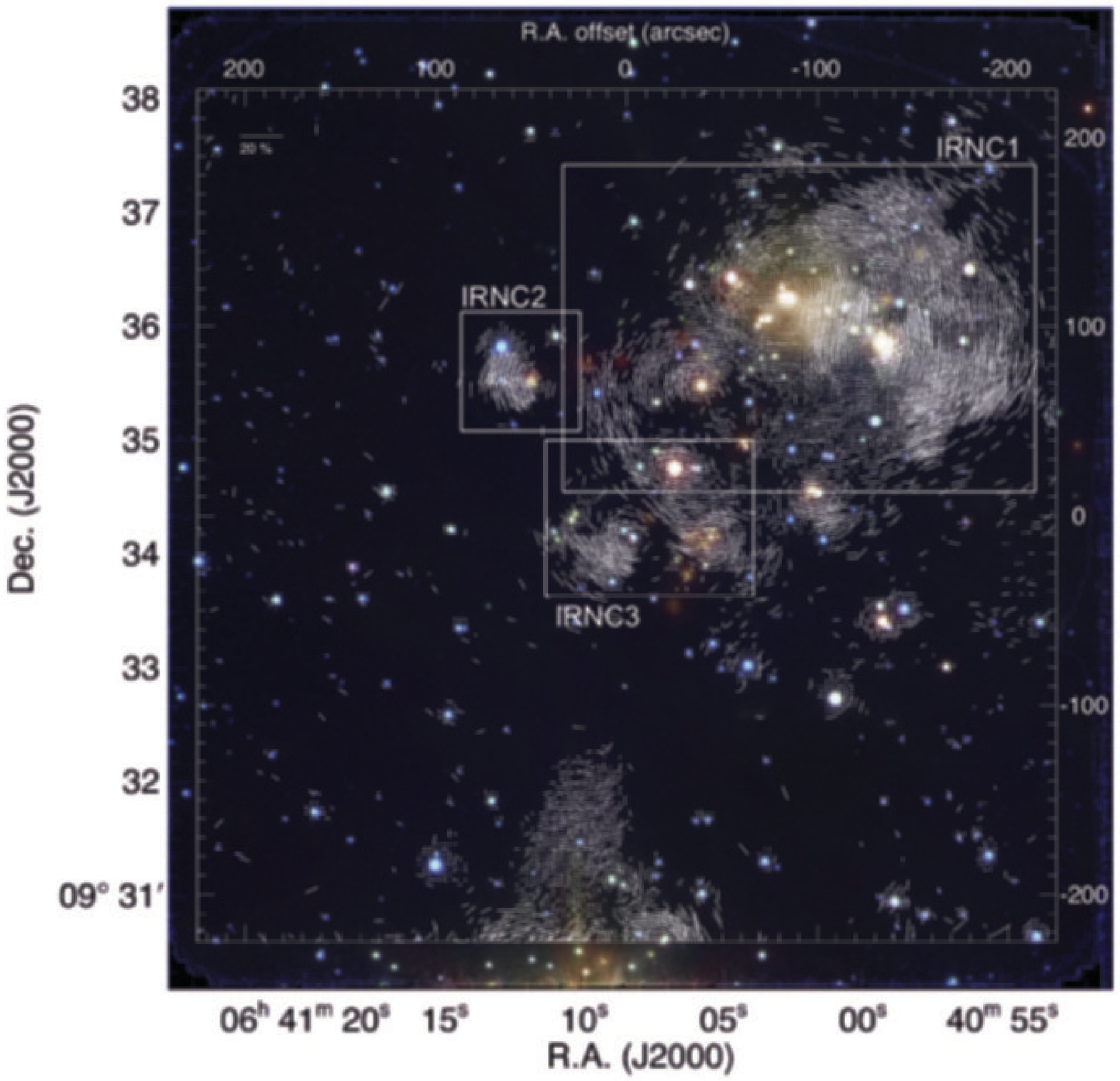}
\vspace{-0.5\baselineskip}
\vspace{-0.5\baselineskip}
\caption{
$H$ polarization vector map
superposed on the color composite image
in the IRSF/SIRPOL $JHK_s$ ({\it blue}),
Spitzer/IRAC 5.8 $\mu$m ({\it green}),
and Spitzer/MIPS 24 $\mu$m ({\it red}) bands.
The length of the vectors is proportional to the degree of polarization.
Shown in the upper left corner is a 20\% vector.
The vectors were made by 6 $\times$ 6 pixel smoothing (twice the seeing size),
and the vectors were shown every 6 pixels 
with intensities greater than 3 $\sigma$ above the mean sky level,
where $\sigma$ is the standard deviation of sky brightness.
Three IRNCs are labeled.}
\end{figure*}

\subsection{Polarimetry of point-like sources}

\subsubsection{Photometry} 

The IRAF DAOPHOT package was used for source detection (Stetson 1987).
The DAOPHOT program automatically detected point-like sources with peak intensities 
greater than 10$\sigma$ above the local sky background, 
where $\sigma$ is the rms uncertainty.
The automatic detection procedure misidentified some spurious sources,
which were removed by careful visual inspection, 
and missed a few real sources such as AR 6B.
We included AR 6B because we were interested in AR 6B (see Section 4.1.1).
Then the IDL photometry package adapted from DAOPHOT was used to perform aperture photometry.
The aperture radius was 3 pixels 
and the sky annulus radius was set to 10 pixels with a 5-pixel width.
The pixel coordinates of the detected sources were matched with the celestial coordinates 
of their counterparts in the 2MASS point source catalog.

The Stokes $I$ intensity of each point-like source was calculated using equation (1).
The magnitude and color of the photometry were transformed into the 2MASS system by
\begin{equation}
   \rm MAG_{2MASS} = MAG_{IRSF} + \alpha_1 \times COLOR_{IRSF} + \beta_1
\end{equation}
and
\begin{equation}
   \rm COLOR_{2MASS} = \alpha_2 \times COLOR_{IRSF} + \beta_2,
\end{equation}
where MAG$_{\rm IRSF}$ is the instrumental magnitude from the IRSF images
and MAG$_{\rm 2MASS}$ is the magnitude from the 2MASS Point Source Catalog.
The parameters were determined by fitting the data
using a robust least absolute deviation method, 
and the mean absolute deviation for each data are 0.11393877, 0.072393811, and 0.084156421
for $J$, $H$, and $K_s$, respectively.
For the magnitudes,
$\alpha_1$ = 0.0484596, $-$0.0307835, and 8.31423e$-$05,   
and $\beta_1$ = $-$4.75512, $-$4.50592, and $-$5.22532 
for $J$, $H$, and $K_s$, respectively.
For the colors,
$\alpha_2$ = 1.066 and 0.989 and $\beta_2$ = $-$0.283 and 0.732
for $J - H$ and $H - K_s$, respectively.
The coefficients $\beta_1$ and $\beta_2$
include both the zero-point correction and aperture correction.
The derived magnitudes are listed in Table 1.
The 10$\sigma$ limiting magnitudes
were 19.2, 18.8, and 16.8 for $J$, $H$, and $K_s$, respectively.

The resulting list contains 314 sources
whose photometric uncertainties
are less than 0.1 mag in all three bands (Table 1).

\subsubsection{Aperture Polarimetry}

Polarimetry of point-like sources (hereafter aperture polarimetry) was carried out
on the combined intensity images for each wave plate angle 
at 10 dithered positions for each set,
instead of using the Stokes $Q$ and $U$ images.
This is because the center of the sources
cannot be determined satisfactorily in the $Q$ and $U$ images.
From the aperture photometry for each wave plate angle image,
the Stokes parameters of each point-like source
were derived by equations (2) and (3).
The aperture and sky radii were the same as those 
used in the aperture photometry of the $I$ images.
The degree of polarization $P$
and the polarization position angle $\theta$
can be calculated by
\begin{equation}
   P_0 = {{\sqrt{Q^2 + U^2}} \over {I}},
\end{equation}
\begin{equation}
   P = \sqrt {P_0^2 - \delta P^2},
\end{equation}
and
\begin{equation}
   \theta = {1\over2} \arctan {U \over Q},
\end{equation}
where $\delta P$ is the uncertainty in $P_0$.
Equation (9) is necessary to de-bias the polarization degree
(Wardle \& Kronberg 1974).
The degree of polarization $P$ was corrected using the polarization efficiencies of SIRPOL.

Table 2 shows the derived source parameters. The uncertainties given in Table 2
(and elsewhere in this paper) are 1$\sigma$ values. 
The aperture polarimetry can indicate the structure of magnetic fields
by measuring dichroic polarization of background stars 
which is produced by aligned interstellar dust grains. 

\section{DISCUSSION}

\subsection{Scattered Radiation Field in the NGC 2264 IRS 2 Region}

There are at least three IRNCs as well as several local IRNe in the outer parts of three IRNCs in Figure 3,
characterized by both a high degree of polarization and a centrosymmetric pattern 
of polarization vectors, albeit often only partial, 
around an illuminating source whose location can often be identified.
Some IRNe overlap with ill-defined boundaries.
The polarization vector patterns around illuminating sources
are often complex, and rarely entirely centrosymmetric.
For example, illuminating sources may be close together,
circumstellar disks and envelopes can involve multiple scattering and dichroic extinction,
and foreground dust in the molecular cloud can also produce dichroic polarization.
That significant regions of centrosymmetric patterns are observed
implies that the amount of foreground dichroic extinction is relatively small, and/or patchy.
In Figure 3, it is clear that each of the protostar cluster members indicated by 24 $\mu$m sources (FTR sources),
is associated with their own local nebulosity,
indicating that each protostar is self-luminous
and associated with the circumstellar matter
such as envelopes or disks.
We discuss the illuminating sources of IRNe in each IRNC in the following subsections.

\subsubsection{IRNC 1}

\begin{figure*}
\epsscale{1.8}
\plotone{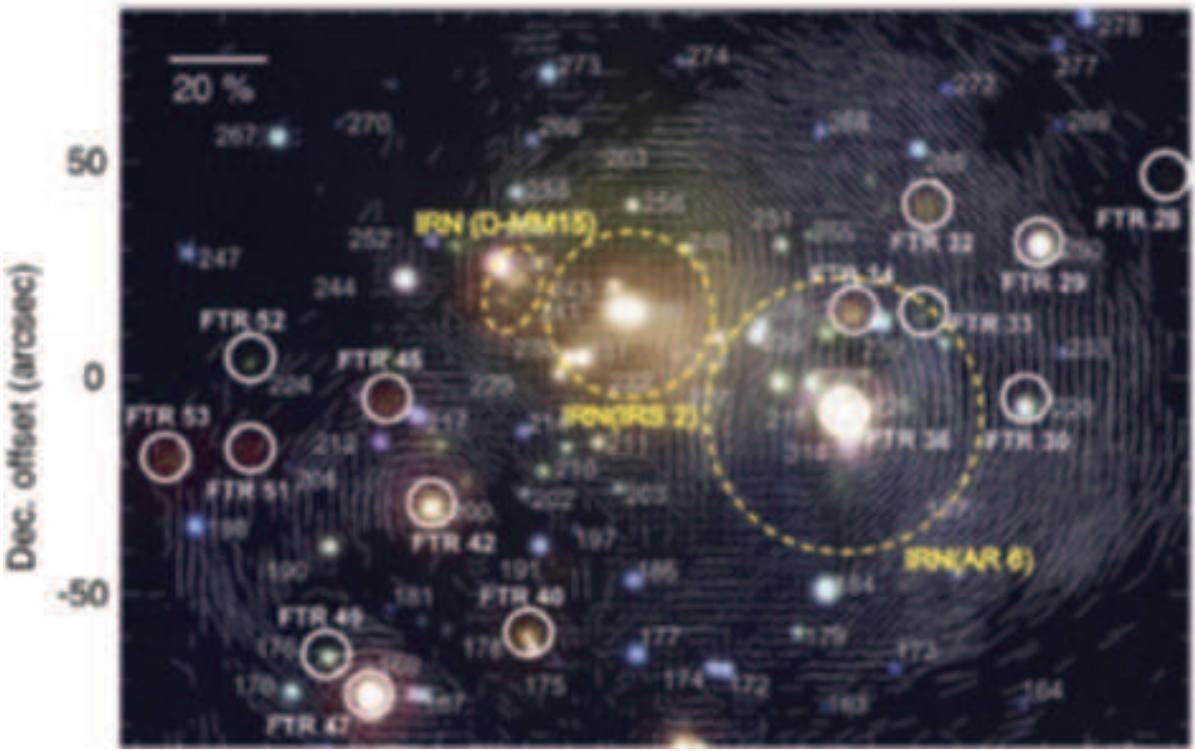}
\vspace{-0.5\baselineskip}
\caption{
$H$ polarization vector map of IRNC 1
superposed on the color composite image
in the IRSF/SIRPOL $JHK_s$ ({\it blue}),
Spitzer/IRAC 5.8 $\mu$m ({\it green}),
and Spitzer/MIPS 24 $\mu$m ({\it red}) bands.
The length of the vectors is proportional to the degree of polarization.
Shown in the upper left corner is a 20\% vector.
The vectors were made by 6 $\times$ 6 pixel smoothing (twice the seeing size).
Detected point-like sources are labeled (Table 1).
24 $\mu$m sources from Forbrich et al. (2010) are denoted by both white circles and labels.
Note that the size of yellow circles that indicate the position of IRNe does not mean the the size of IRNe.}
\end{figure*}

\begin{figure*}
\epsscale{1.9}
\plotone{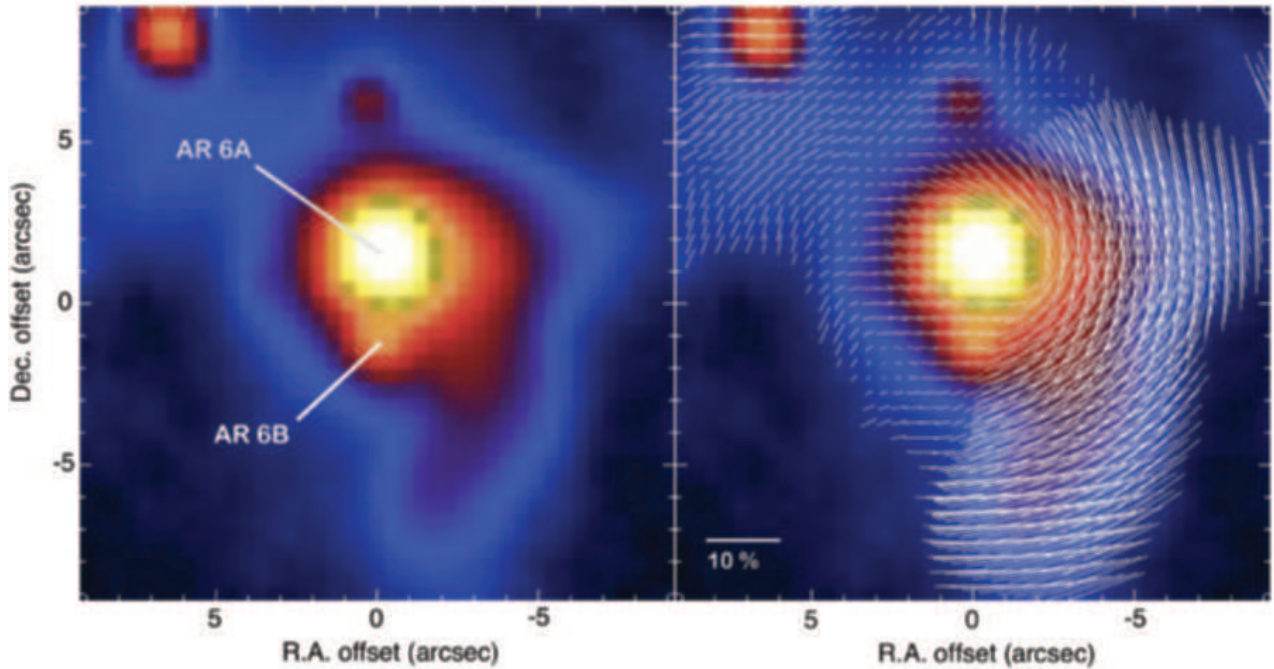}
\vspace{-0.5\baselineskip}
\vspace{-0.5\baselineskip}
\caption{
(left panel) Stokes $I$ image of IRN (AR 6) in the $H$ band.
AR 6A and AR 6B are labeled.
(right panel) $H$ polarization vector map of IRN (AR 6).
Shown in the bottom left corner is a 10 \% vector.
The vectors were made by 3 $\times$ 3 pixel smoothing (the seeing size).}
\end{figure*}

Figure 4 shows a close-up of the $H$ polarization vector map 
of IRNC 1 shown in Figure 3. 
It is clear that the IRNC 1 region is dominated by 
various highly polarized local IRNe with illuminating sources, 
while the outer parts are roughly centrosymmetric around the central stellar sources (Figure 4).

The main illuminating source showing a large, centrosymmetric nebulosity 
extending to the south-west (more than 1$\arcmin$) 
and also probably to the north-east (up to the middle point between AR 6 and IRS 2) 
is most likely to be AR 6. 
AR 6 consists of three stellar components, 
the brightest star AR 6A, 
the southern source 6B,
and the faintest companion 6C.
Unfortunately, our data could not resolve AR 6C 
because the distance between AR 6A and AR 6C is only 0\farcs84,
smaller than our seeing size.
AR 6A$+$6C and AR 6B correspond to our source 218 and 314, respectively.
The polarization of source 218 is 5--7\% at $\sim$74$\arcdeg$,
and the polarization of source 314 is 2--5\% at 42$\arcdeg$--74$\arcdeg$.

The large IRN clearly shows a centrosymmetric pattern, 
indicating that the nebulosity is most likely associated with the central source AR 6. 
Therefore, we call it IRN (AR 6).
This nebula corresponds to the blue region surrounding AR 6 in Figure 1, 
suggesting the central source is ``evolved".
It is consistent with the illuminating source AR 6 being a visible YSO.

As shown in Figure 5, an enlargement of IRN (AR 6), 
AR 6 itself shows a compact ($\sim$20$\arcsec$ $\times$ 10$\arcsec$) cometary nebulosity, 
extending from the west of the stars and curving around them to the south 
(Aspin \& Reipurth 2003). 
To the east of AR 6, 
polarizations are partly offset and not always centrosymmetric around AR 6 
but are directing south-east of AR 6. 
There may be a polarization disk (Bastien \& M\`{e}nard 1988, 1990; Tamura et al. 1991) 
around AR 6.
However, it is rare that a polarization disk is observed around a visible YSO 
because the polarization disk is believed to be caused by multiple scattering in the dense disk. 
Aspin \& Reipurth (2003) suggested that 
AR 6A is a young star within the NGC 2264, 
as an FUor-like pre-main-sequence object, 
and all of the nebulosity seen may be connected to AR 6A. 
AR 6B is also in an elevated FUor state as an FUor-like object, 
but it does not appear intimately associated with the nebulosity.

IRS 2 also belongs to the infrared reflection nebula cluster IRNC 1. 
It shows a generally centrosymmetric pattern, 
even though there are slight changes in the vector pattern 
at several parts of the nebula. 
The IRN (IRS 2) is approximately centrosymmetric out to $\sim$30$\arcsec$,
except for around point-sources in Figure 4.
In the outer part of up to 1\farcm5 to the north-west, 
the nebula also seems to be illuminated by IRS 2 rather than AR 6. 
Note that near the middle point between IRS 2 and AR 6, 
the polarization pattern is a mixture of the two polarization patterns.

IRS 2 is resolved into two sources in our images 
(RNO-E and RNO-W; Castelaz \& Grasdalen 1988). 
RNO-E and RNO-W correspond to our sources 241 and 240, respectively (see Table 1).
It is noteworthy that 
the polarization pattern very close to these sources (a few arcsec) 
is not centrosymmetric but rather aligned in IRN (IRS 2); 
this can trace the magnetic fields rather than the local nebula.

There are other notable reflection nebulae and 
their illuminating sources in this field 
besides those associated with AR 6 and IRS 2 as described below.

The most interesting feature of Figure 4 is the small nebula, IRN (D-MM15),
associated with a faint source or a cluster of faint sources $\sim$30$\arcsec$ 
to the east of IRS 2. 
This does not follow the centrosymmetric pattern around IRS 2 
or the nearest source 246, 
suggesting that this faint source is an independent self-luminous source. 
This source is not point-like and is only visible in the $K_s$ band. 
It appears that this source corresponds to 
a millimeter/submillimeter source D-MM15 of Peretto et al. (2006). 
The position of D-MM15 is
$\alpha$ = 06$^h$ 41$^m$ 04$\fs$6, $\delta$ = +09$\arcdeg$ 36$\arcmin$ 19$\arcsec$ [J2000.0].
Hashimoto et al. (2008) proposed an evolutionary sequence of massive YSOs 
based on the morphology of the associated IRNe. 
The Type A IRNe defined by Hashimoto et al. (2008) 
have a non-point-like morphology
and extend in a direction relatively perpendicular to the polarization vectors,
and is similar to IRN (D-MM15) as shown in Figure 4.
They suggested that the monopolar or bipolar IRN of aligned vectors
corresponds to the earlier phase 
because multiple scattering is dominant in the circumstellar matter 
around the youngest YSOs.

Source 200 is within the red nebula around IRS 2, and
the vector pattern is partly centrosymmetric in the outer part of source 200.
However, the vectors in the inner part are well aligned in one direction
that is associated with the magnetic field direction in the IRS 2 region.
It is highly polarized (P = 4--9\%) in all three bands (Table 2).

Source 169 shows similar vector patterns to source 200.
The vector pattern is partly centrosymmetric in the outer part of source 169,
and it is also highly polarized (P = 5--10\%) in all three bands (Table 2).

\subsubsection{IRNC 2}

A faint red nebula around FTR 51 and FTR 53 (sources 51 and 53 of FTR) 
extends to the east of source 200 in Figure 4.
The vector pattern of this nebula is not centrosymmetric around source 198, 
suggesting that the scattered radiation field here is dominated
by the sources in IRNC 1.
However, there is another IR cluster to the east 
whose nebulae are distinct from IRNC 1.

Figure 6 shows the polarization vector maps around FTR 56,
as a close-up of the $H$ polarization vector map of IRNC 2 shown in Figure 3.
This source is very red and in fact not point-like. 
The polarization map shows that 
the source itself is a reflection nebula 
and it extends significantly to the east (up to $\sim$20\arcsec) 
but is partly interrupted by the blue point source 201. 
A small ($\sim$10\arcsec) reflection nebula extending to the west is also clearly recognized. 
There is an independent nebula around the north-east blue sources 216 and 213
but the polarization pattern is not centrosymmetric. 
The north-west source 223 is also associated with a nebula 
but again its polarization pattern is not simple. 
The bright 24 $\mu$m sources FTR 51 and FTR 53
are totally invisible in the near-IR.

\subsubsection{IRNC 3}

\begin{figure}[!t]
\epsscale{1.0}
\plotone{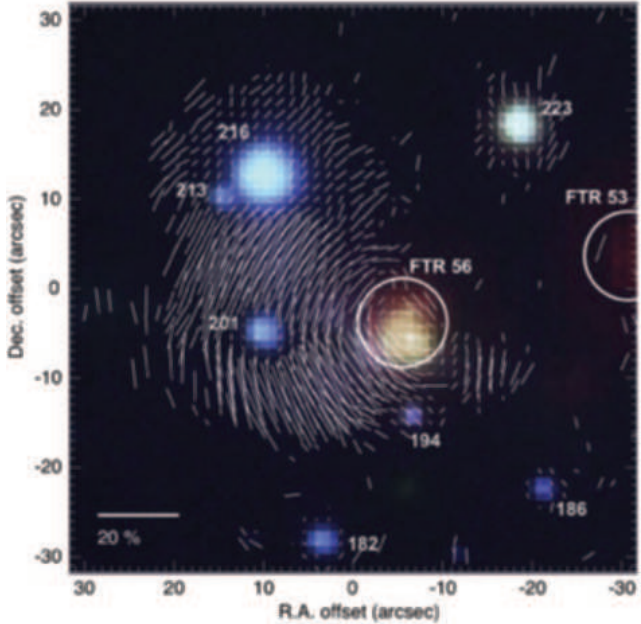}
\vspace{-0.5\baselineskip}
\vspace{-0.5\baselineskip}
\caption{
Same as Figure 4 for IRNC 2.}
\end{figure}

\begin{figure*}
\epsscale{1.9}
\plotone{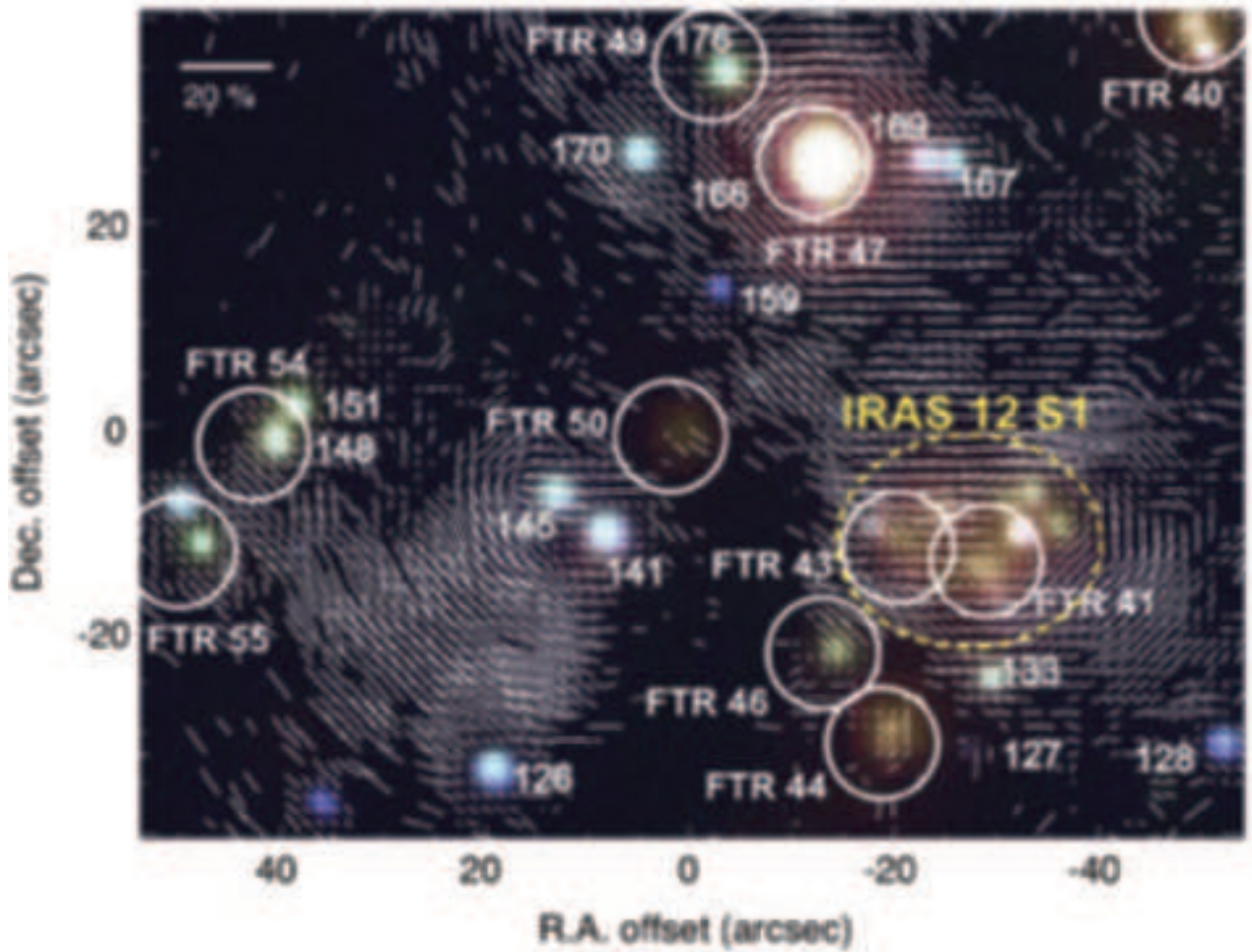}
\vspace{-0.5\baselineskip}
\vspace{-0.5\baselineskip}
\caption{
Same as Figure 4 for IRNC 3.}
\end{figure*}


Figure 7 shows a close-up of the $H$ polarization vector map of IRNC 3 shown in Figure 3,
including the IRAS 12 S1.
As shown in Figure 7, 
the vector pattern around IRAS 12 S1 itself is not highly polarized  ($\sim$10 \% or less) 
compared with those in IRNC 1 and IRNC 2 ($\sim$10 \% or more), 
but our data suggest a clear association of a medium-size reflection nebula 
with a centrosymmetric pattern around this ``microcluster'' of protostars 
(Teixeira et al. 2006; YTL). 
In the outer part (up to $\sim$20$\arcsec$), 
the nebula is roughly centrosymmetric, 
in particular to the south, south-west, and north-west in the $H$ band. 
There is a more simple, interesting pattern in the $K_s$ band.
However, the central part and the region to the north-east are very complex.

\begin{figure*}
\epsscale{1.7}
\plotone{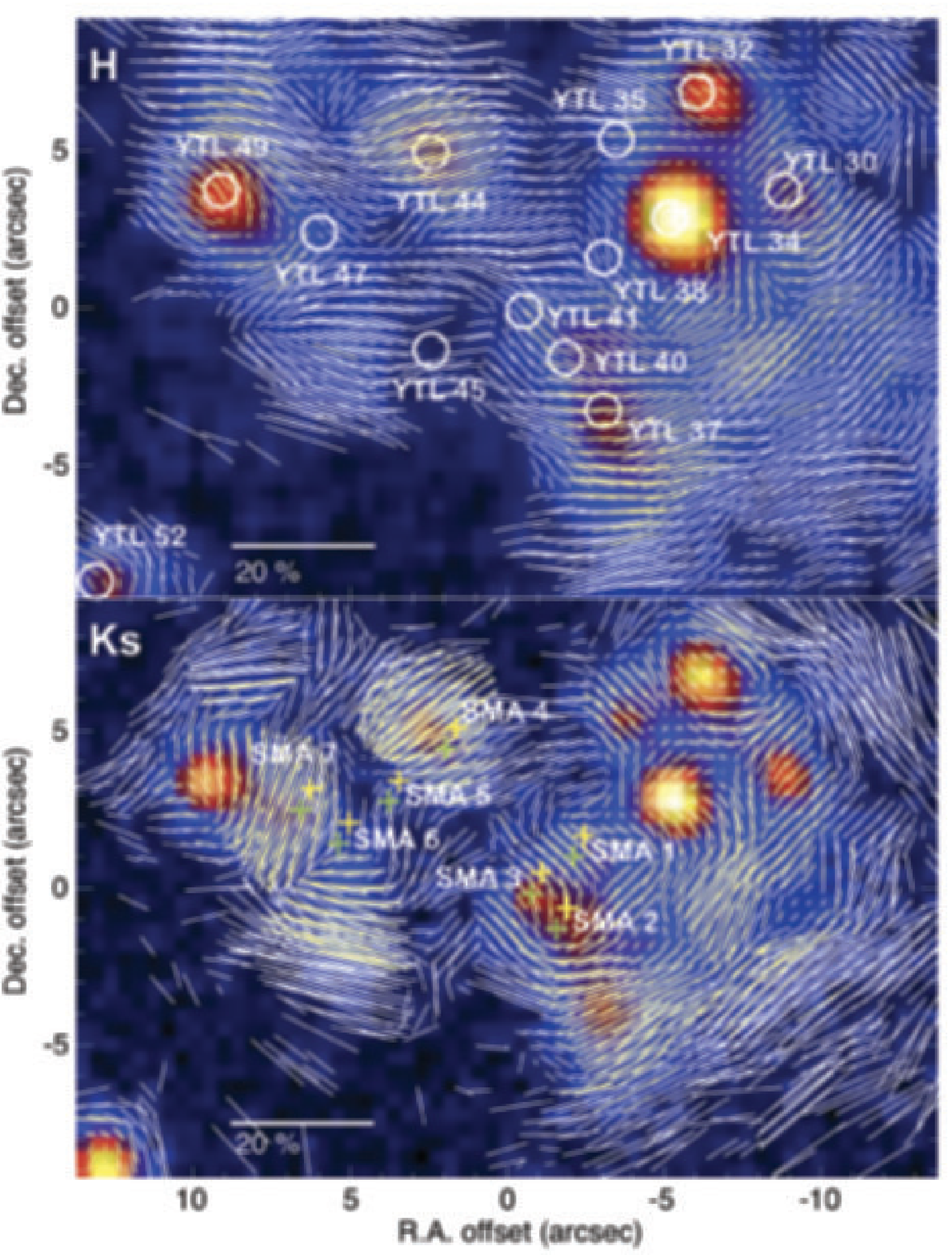}
\vspace{-0.5\baselineskip}
\vspace{-0.5\baselineskip}
\caption{
$H$ and $K_s$ polarization vector maps around the IRAS 12 S1 core of IRNC 3 superposed on each Stokes $I$ image.
The vectors were made by 3 $\times$ 3 pixel smoothing (the seeing size),
and shown every 1 pixels with intensities greater than 3$\sigma$ and 5$\sigma$ 
above the mean sky level in the $H$ and $K_s$ bands, respectively.
Additionally, the yellow vectors were drawn for 
greater than 6$\sigma$ and 10$\sigma$ in the $H$ and $K_s$ bands, respectively.
(top panel)
$H$ polarization vector map.
The circles mark the positions of the sources
detected at the $K$ band with Baade/PANIC
with a stellar FWHM of 0\farcs37--0\farcs44 (Young et al. 2006).
(bottom panel)
$K_s$ polarization vector map. The seven plus signs mark
the positions of the SMA continuum data (Teixeira et al. 2007).
Yellow plus signs: Original positions corresponding with Figure 1 of Teixeira et al. (2007).
Green plus signs: Shifted positions to match their counterparts.}
\end{figure*}

Figure 8 shows enlargements of the $H$ and $K_s$ polarization vector maps for the IRAS 12 S1,
i.e., in the right lower quadrant of Figure 7. 
These vector maps show very complex vector patterns slightly different from each other 
because of the sources only detected in the $K_s$ band, 
or the difference between the amount of foreground dichroic extinction by different wavelengths.
It also indicates that there might be other hidden sources 
as well as infrared sources and submillimeter sources in Figure 8.

In the $H$ polarization vector map, 
the positions of the sources detected in the $K_s$ band with Baade/PANIC (YTL) 
are marked with circles and their source numbers.
YTL reported several knots of molecular hydrogen in this ``microcluster'', 
for example, between YTL 47 and YTL 49 and between YTL 40 and YTL 37.
They also showed that YTL 44, associated with molecular hydrogen emission,
has a distinctly bipolar appearance with an opaque dust lane,
most likely caused by a circumstellar disk.
Of particular interest is YTL 44,
which shows a monopolar IRN of aligned vectors (Type A; Hashimoto et al. 2008).
In the $K_s$ polarization vector map,
the positions of the seven SMA continuum data 
using the Submillimeter Array whose synthesized beam had dimensions 
of 1$\farcs$4 $\times$ 1$\farcs$3 (Teixeira et al. 2007)
are marked with plus signs 
on the basis of the source positions from their $K_s$ band image (0$\farcs$125 pixel$^{-1}$).
Teixeira et al. (2007) reported that SMA 1 and SMA 3 have mid-IR counterparts
and SMA 2, SMA 6, and SMA 7 are associated with diffuse mid-IR emission.
SMA 4 and SMA 5 do not appear to have any near- and/or mid-IR emission,
even though there is near- and mid-IR diffuse emission located  between these sources.
In Figure 8 (bottom panel),
the position of the SMA sources are slightly shifted (by $\sim$1$\arcsec$),
which is within the positional errors of the submillimeter array.
We found that the shifted position show a better correlation 
between the near-IR nebula and the submillimeter sources.

FTR 50 and FTR 44 was classified as ``genuine protostars'' based on detailed SEDs (FTR).
FTR 41 and FTR 43 that are multiple sources in IRAS 12 S1 also clearly showed protostar SEDs (FTR).
They all do not show a centrosymmetric pattern of polarization vectors in Figure 7.

\subsubsection{Local IRNe}

We found that several 24 $\mu$m sources such as FTR 39, FTR 38, FTR 37, and FTR 35 
are associated with a local IRNe outside of three IRNCs in the observed region.
The vector patterns around these local IRNe except for FTR 39 were not highly polarized 
compared with those in three IRNCs.
IRN (FTR 39) is considered as one of members of IRNC 1.

\subsection{Multiple Magnetic Fields in the IRS 2 Region}

\begin{figure}[!t]
\epsscale{1.0}
\plotone{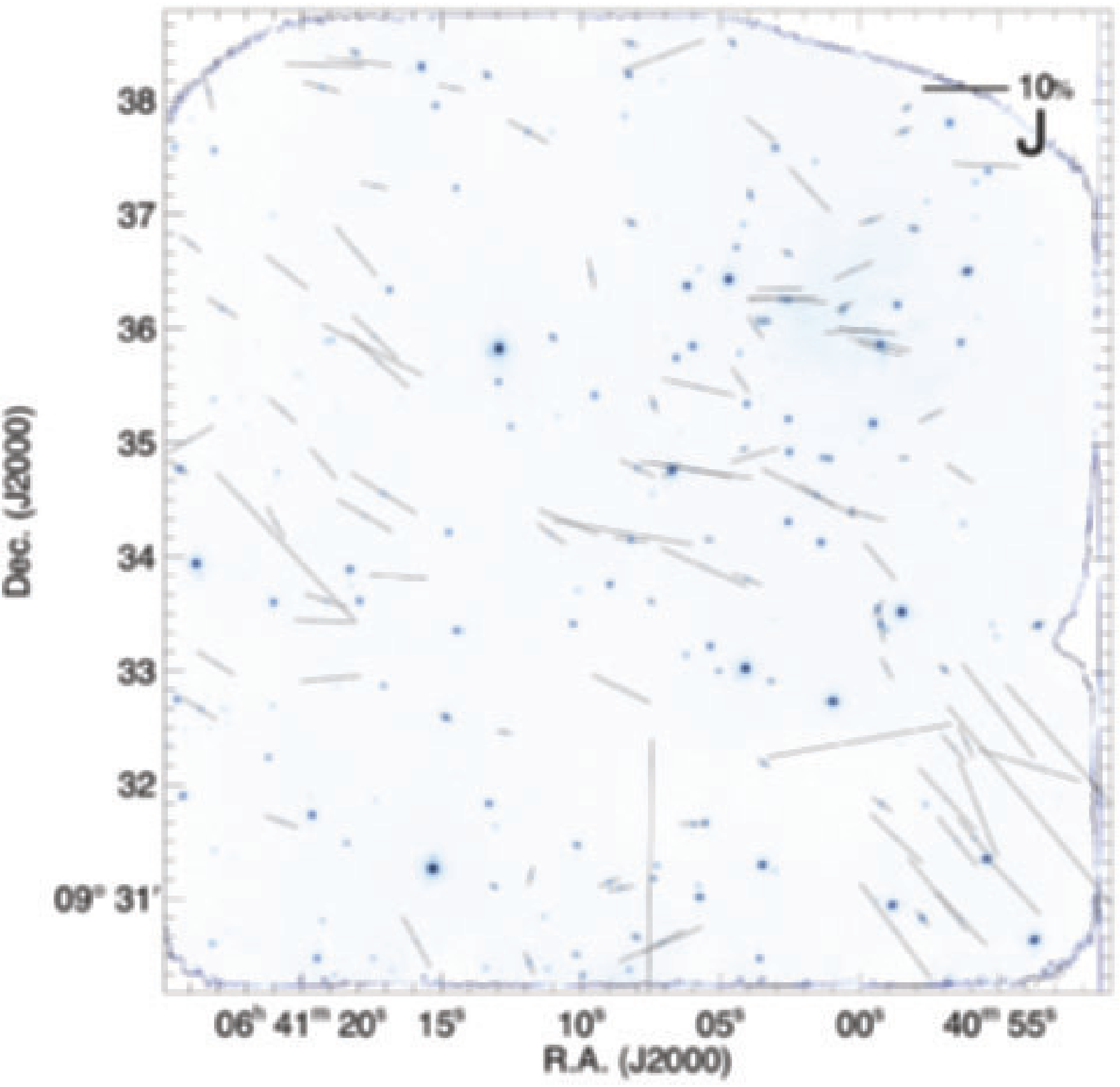}
\caption{
Stokes $I$ image of the $J$ band with polarization vectors for point-like sources 
having $P/\delta P \ge$ 3.
The length of the vectors is proportional to the degree of polarization.
Shown in the upper right corner is a 10\% vector.
Note that there are bad pixel clusters
around the upper-left and upper-right corners
and the middle of the right boundary.}
\end{figure}

\begin{figure}[!t]
\epsscale{1.0}
\vspace{-0.9\baselineskip}
\plotone{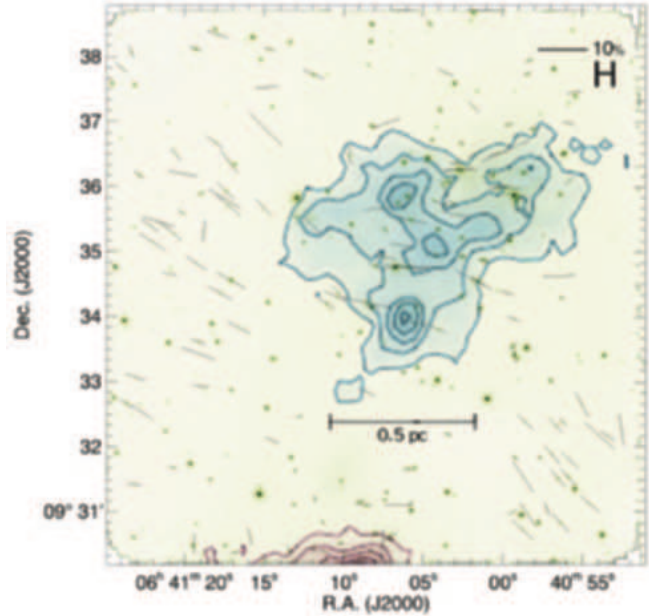}
\caption{
Same as Figure 9 for the $H$ band with contours of 870 $\mu$m continuum emission 
(Williams \& Garland 2002). Contours are at 0.5, 1.0, 1.5, 2.0, and 2.5 Jy beam$^{-1}$.
Red contours: dense cluster of IRS 1 region.
Blue contours: dense cluster of IRS 2 region.}
\end{figure}

\begin{figure}[!t]
\epsscale{1.0}
\plotone{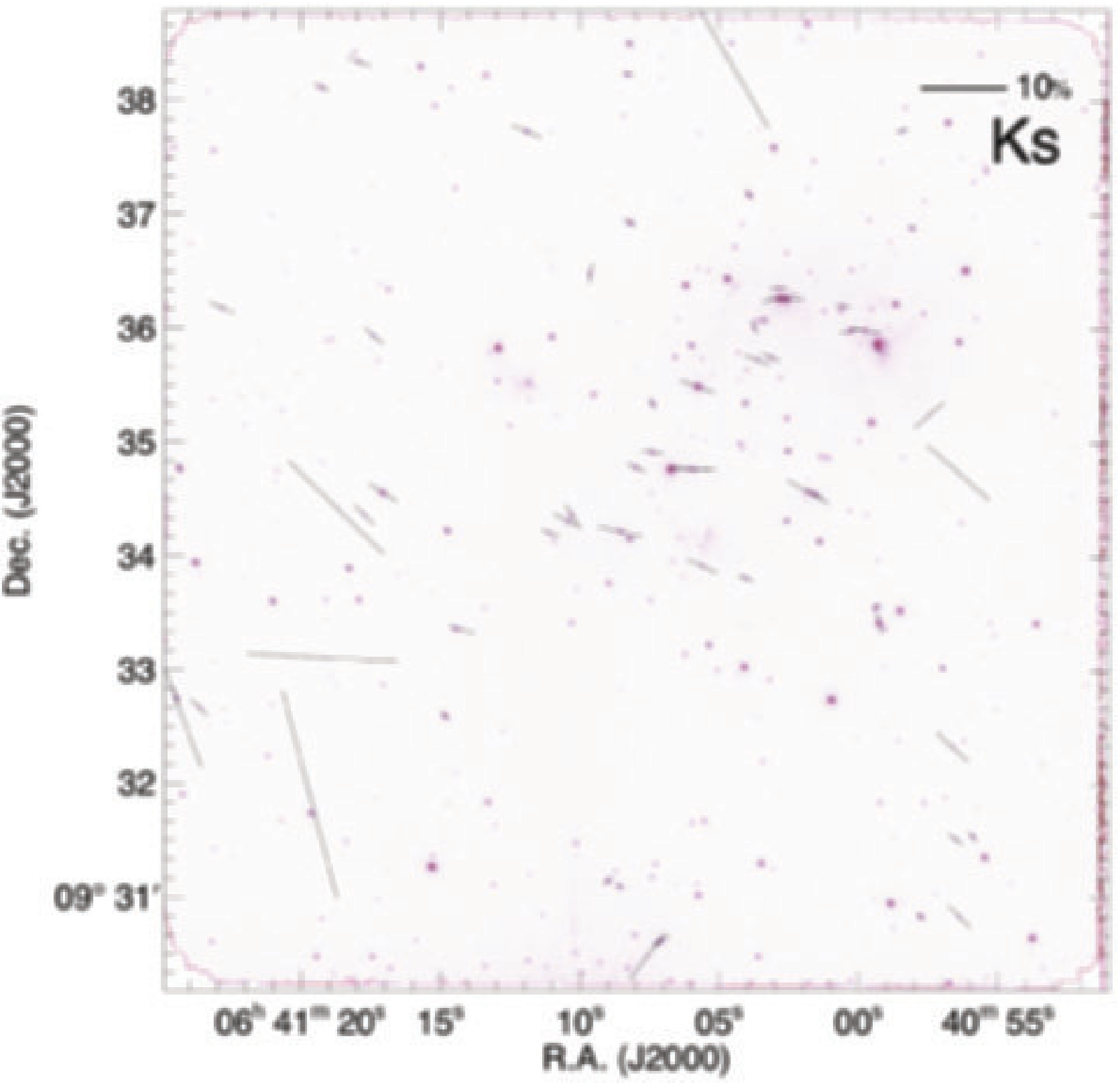}
\caption{
Same as Figure 9 for the $K_s$ band.}
\end{figure}

Figures 9--11 show the polarization maps of point-like sources 
superposed on the Stokes $I$ images.
The aperture polarimetry of the 314 point-like sources
resulted in the positive detection ($P/\delta P \ge$ 3) of 181, 212, and 97 sources
in the $J$, $H$, and $K_s$ bands, respectively, 
and 241 sources in at least one of the three bands.

The relation between the polarimetric and spectral data
may be useful in understanding the nature of the polarization.
The degree of polarization appears to be correlated with near-IR colors
(Figure 12).
The empirical relation for the upper limit of interstellar polarization
suggested by Jones (1989) is
\begin{equation}
   P_{K, \rm max} = \tanh \left\{ 1.5 E(H-K) {{1-\eta}\over{1+\eta}} \right\},
\end{equation}
where $\eta$ = 0.875 and $E(H-K)$ is the reddening arising from extinction.
As shown in Figure 12,
most of the sources are within this limit.
A few sources are above the $P_{\rm max}$ limit,
but their uncertainties are large.
The near-IR polarization-to-extinction efficiency of the point-like sources in the IRS 2 field 
is consistent with that caused by aligned dust grains in the dense interstellar medium.
Therefore, the near-IR polarizations of these sources 
are most likely dominated by dichroic extinction from aligned dust grains by a magnetic field,
and the intrinsic polarization, if any, 
does not significantly affect the observed degree of polarization.
However, this result does not completely exclude the possibility that some of the sources
have intrinsic polarization because depolarization is also possible.

Aperture polarimetry in the $K$ band using a 4$\arcsec$ aperture is reported by
Castelaz \& Grasdalen (1988), for both RNO-E and RNO-W.
The polarization degree of RNO-E was 3.3 $\pm$ 0.3 \%
with a position angle of 30$\arcdeg$ $\pm$ 3$\arcdeg$,
and the polarization degree of RNO-W was 3.5 $\pm$ 0.3 \%
with a position angle of 7$\arcdeg$ $\pm$ 3$\arcdeg$.
In our near-IR data,
the polarization degree of RNO-E 
is 4--8\% with a position angle of 91$\arcdeg$--98$\arcdeg$ with smaller errors.
The polarization position angle of Castelaz \& Grasdalen (1988)
is somewhat smaller than our near-IR measurements.
It is not clear what caused the difference in these sources.

\subsubsection{Source Classification}

The magnetic field structure of molecular clouds can be inferred
from the polarization produced by dichroic extinction,
provided sources are selected with no intrinsic polarization.
YSOs in the cloud can exhibit a substantial degree of intrinsic polarization
caused by circumstellar material.
Such sources may show a large IR excess 
which can be identified from multiwavelength photometry.

Figure 13 shows a color--color diagram for all sources detected in all three bands.
The diagram is divided into several domains. 
Based on the location in this diagram,
sources can be classified into a few groups (Lada \& Adams 1992).
The area near the locus of main-sequence/giant stars is called domain A0.
Sources in domain A0 are either field stars (dwarfs and giants)
or pre-main-sequence (PMS) stars with little IR excess
(weak-lined T Tauri stars and some classical T Tauri stars)
and with little reddening.
There is a clear gap just above domain A0,
and the area above this gap in the direction of the reddening vector
is called domain Ar.
Sources in domain Ar are either field stars or PMS stars
with little IR excess and with substantial reddening.
Domain B is the area next to domain Ar
in the direction of higher $H$ -- $K_s$ (to the right) 
and above the locus of classical T Tauri stars.
Sources in domain B are PMS stars with IR excess emission from disks.
Domain C is the area next to and to the right of domain B.
Sources in domain C are IR protostars or Class I sources.
Herbig AeBe stars tend to occupy lower parts of domains B and C.
This classification based on the color--color diagram, however,
is far from perfect.
A certain fraction of classical T Tauri stars 
may reside in domains A0 or Ar,
some protostars can be found in domain B,
and some extremely reddened AeBe stars may be found among protostars
(Lada \& Adams 1992).
This ``contamination'' will eventually contribute to the uncertainty
in statistical quantities
derived from the classification,
but the estimation of this uncertainty is beyond the scope of this paper.

There are 125 sources in domain A0,
and they are collectively called group A0.
They are either foreground stars 
or those seen along the line of sight with little extinction.
One hundred fifty one sources were found in domain Ar.
These sources (group Ar) are either background stars or PMS stars
in the Monoceros OB1 cloud.
They are the most useful sources for the study of the magnetic fields
in the cloud (see Section 4.4 of Kwon et al. 2010).

\begin{figure*}
\epsscale{1.9}
\plotone{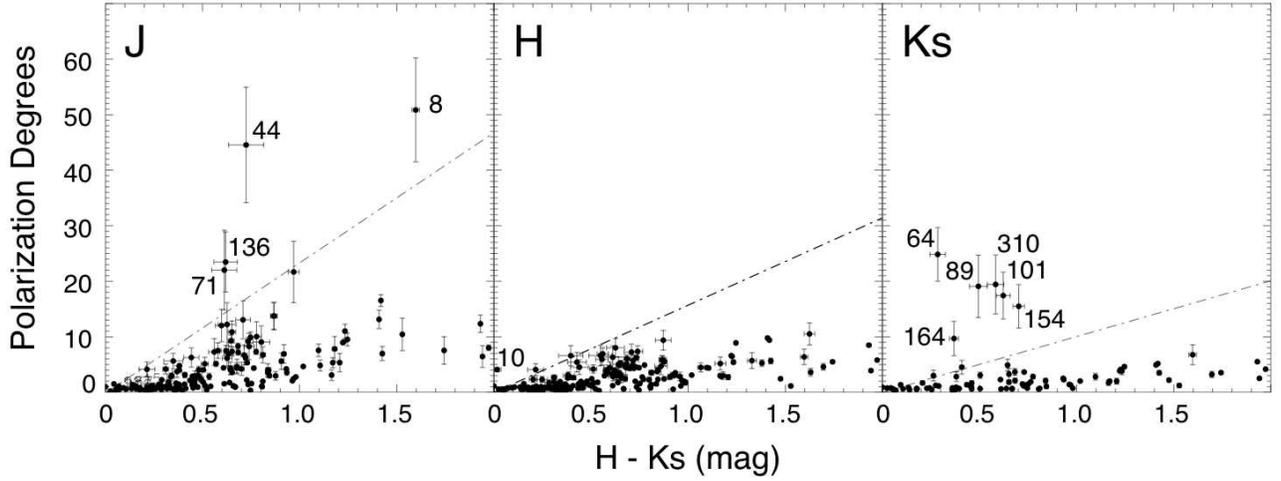}
\vspace{-0.5\baselineskip}
\vspace{-0.5\baselineskip}
\caption{
Degree of polarization vs. $H - K_s$ color.
Several sources with large uncertainties are labeled (Table 2).
Dot-dashed lines:
empirical upper limits ($P_{\rm max}$; Jones 1989).}
\end{figure*}

\begin{figure}[!t]
\epsscale{0.8}
\plotone{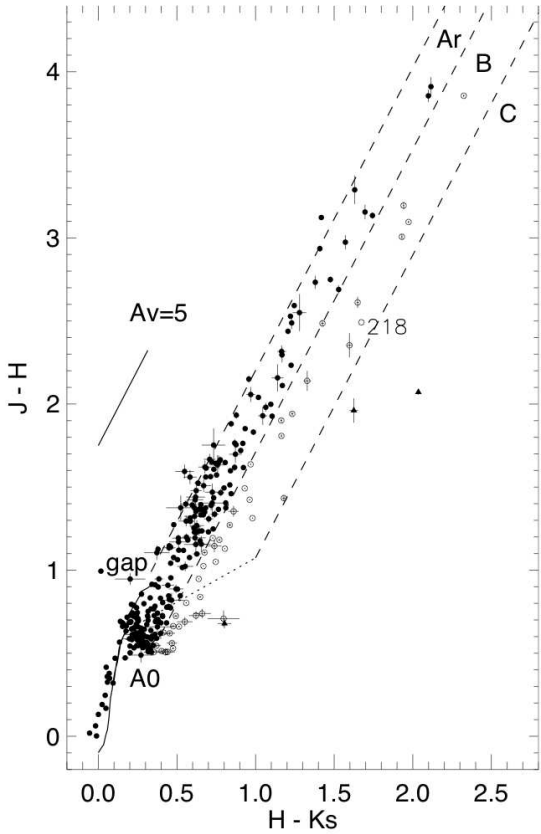}
\vspace{-0.5\baselineskip}
\vspace{-0.5\baselineskip}
\caption{
Color--color diagram of the point-like sources in the IRAS 12 S1 field.
Filled circles: group A0 and Ar sources.
Open circles: group B sources.
Open triangles: group C sources.
Solid curve:
locus of main sequence and giant branch stars (Bessell \& Brett 1988).
Dotted line:
locus of classical T Tauri stars (Meyer et al. 1997).
Dashed lines:
boundaries between domains A, B, and C (see Section 4.2.2).
Solid line:
reddening vector.}
\end{figure}

Forty-five sources were found in domain B.
These sources (group B) may be PMS stars
associated with the Monoceros OB1 cloud.

Three sources (sources 255, 307, and 314) are located in domain C.
AR 6B (source 314) is considered a variable star of FU Ori type 
(Aspin \& Reipurth 2003).

\subsubsection{Magnetic Field Strength}

The sources in group Ar (Figure 13) are best for studying the magnetic fields in the molecular cloud 
because they are likely to have little intrinsic polarization, 
and the high reddening can lead to polarization through dichroic extinction
that is selective attenuation of different components of the electric vector 
when light passes through a medium
in which the grains are aligned by a magnetic field.
Since there is significant difference
between the vectors of the cluster core and outer parts in Figures 9--11,
only the sources in the cluster core were chosen 
to estimate the magnetic field strength of the IRS 2 region, 
among group Ar sources.
In Figure 10, contours of 870 $\mu$m continuum emission (Williams \& Garland 2002) were superposed
on the $H$ polarization vector map. 
In the following discussion, 
we designate the sources within blue contours (dense cluster of IRS 2 region) as that of cluster core, 
and the sources outside blue and red contours as that of the perimeter of the cluster core. 
Note that the sources within red contours (dense cluster of IRS 1 region) are excluded 
from following discussions for magnetic fields.

\begin{figure}[!t]
\epsscale{1.0}
\plotone{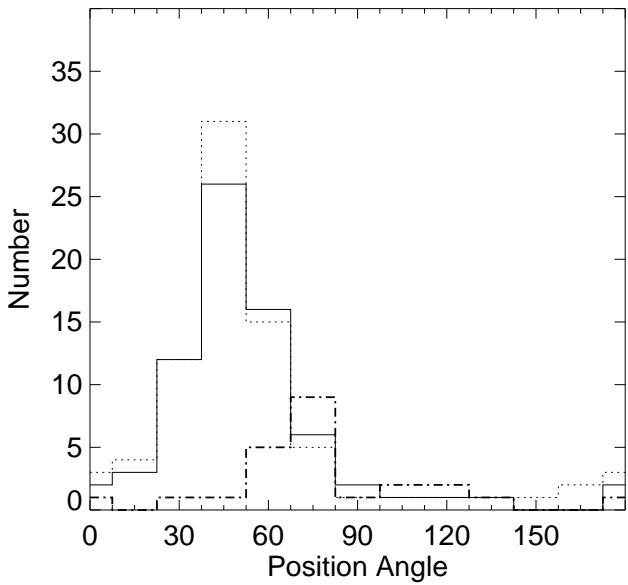}
\vspace{-0.5\baselineskip}
\vspace{-0.5\baselineskip}
\caption{
Histograms for the $H$ polarization position angles of the group Ar sources
in both the cluster core and perimeter. 
Note that 15$\arcdeg$ bin is used because the sample size is small.
For the point-like sources of the cluster core, the mean angle is 81$\arcdeg$
and for the sources in the perimeter of the cluster core the mean angle is 58$\arcdeg$.
The mean angle is 60$\arcdeg$.
Dot-dashed line:
histogram of polarization position angles for the group Ar sources having $P/\delta P \ge$ 3 
in the cluster core, from our near-IR data.
Solid line:
histogram of polarization position angles for the group Ar sources having $P/\delta P \ge$ 3 
in the perimeter of the cluster core, from our near-IR data.
Note that for the point-like sources of the dense cluster
the distribution is relatively widespread,
but the sample size is too small to discuss in detail.
Also shown is the histogram of polarization position angles at a wavelength of 350 $\mu$m.
The distribution of the polarization position angles at 350 $\mu$m is in better agreement
with that of the $H$ band in the perimeter of the cluster core than in the cluster core.
Dotted line:
histogram of polarization position angles of all the 77 sources in IRS 2 region
obtained from 350-$\mu$m observations (Table 2 of Dotson et al. 2010)
with the criterion $F > 3\sigma_F$, where $F$ is the total unpolarized flux density.}
\end{figure}

Figure 14 shows the histogram 
for the $H$ polarization position angles of the group Ar sources 
in both the cluster core and perimeter.
The dispersion of the polarization position angles toward the IRAS 12 S1 core,
covering a region of about a 4$\arcmin$ $\times$ 4$\arcmin$,
is larger than that of the point-like sources in the perimeter of the cluster core.
For the point-like sources of the dense cluster, 
the mean angle is 81$\arcdeg$ with a standard deviation of 29$\arcdeg$.
For the sources in the perimeter of the cluster core, 
the mean angle is 58$\arcdeg$ with a standard deviation of 24$\arcdeg$.
The difference between the cluster core and the perimeter is 23$\arcdeg$, 
which is smaller than the standard deviation of each of them.
Note that there is a systematic gradient in the magnetic field orientation 
over the imaged field, 
even though there is relatively large dispersion of polarization angles 
in the cluster core.

Although the polarization measurement does not provide 
a direct estimate of the magnetic field strength 
at each data point in the image, 
a rough estimate over a large region is possible 
by statistically comparing the dispersion of the polarization orientation 
with the degree of turbulence in the cloud (Chandrasekhar \& Fermi 1953). 
Assuming that velocity perturbations are isotropic, 
the strength of the magnetic field projected on the plane of the sky 
can be calculated by
\begin{equation}
   B_p = {\cal Q} \sqrt{4\pi\rho}\ {{\delta v_{\rm los}}\over{\delta\theta}},
\end{equation}
where ${\cal Q}$ is a factor to account for various averaging effects,
$\rho$ is the mean density of the cloud,
$\delta v_{\rm los}$ is the rms line-of-sight velocity,
and $\delta\theta$ is the dispersion of the polarization angles.
Ostriker et al. (2001) suggested that ${\cal Q}$ $\sim$0.5
is a good approximation
when the angle dispersion is small ($\delta\theta \lesssim$ 25\arcdeg)
from numerical simulations,
and Houde (2004) also suggested that
a correction factor of $\sim$0.5 is appropriate in most cases
when the magnetic field is not too weak.
Since the magnetic field seems to be ordered over the IRS 2 region,
the magnetic field is expected to be strong 
and we adopt a correction factor of 0.5 to estimate the magnetic field strength,
even though the angle dispersion (29\arcdeg) in the cluster core 
is a little larger than the 25$\arcdeg$ suggested by Ostriker et al. (2001).
From observations of cluster forming clumps in the N$_2$H$^+$ (1--0),
Peretto et al. (2006) estimated a mean H$_2$ density of $6.0 \times 10^4$ cm$^{-3}$
assuming a spherical volume and a mean clump diameter of 0.9 pc. 
They also estimated a 3D velocity dispersion of 1.4 km s$^{-1}$, 
assuming isotropic motion. 
Then, from equation (12), the strength of the magnetic field projected on the plane of the sky 
is derived to be $B_p \approx$ 100 $\mu$G.
The uncertainty in this estimate is rather large
because the observed IRS 2 field is only a part of the Monoceros OB1 cloud and 
because the mean density $\rho$ used in above equation is small 
due to imperfect coupling between ions and neutrals.
It has been suggested that the gas densities are too low
toward the edge of the CO outflow in the IRS 2 region (Wolf-Chase \& Walker 1995).
Thus, it should be taken as an order-of-magnitude estimate.
The estimated magnetic field strength of the IRS 2 region
is similar to that of other molecular clouds (20--200 $\mu$G)
derived using the Chandrasekhar--Fermi method
(e.g., Andersson \& Potter 2005; Poidevin \& Bastien 2006;
Alves et al. 2008; Kwon et al. 2010; Sugitani et al. 2010).

\subsubsection{The Role of Magnetic Fields in Cluster Formation}

\subsubsubsection{4.2.3.1 Galactic vs. Local-cloud Magnetic Fields}

Previous observations suggest that
the local magnetic field associated with molecular clouds 
correlates with the Galactic magnetic field
(Kobayashi et al. 1978; Dyck \& Lonsdale 1979).
This indicates that the Galactic magnetic field is likely to play an important role
in the formation and evolution of molecular clouds.
For example, Dyck \& Lonsdale (1979) compared infrared aperture polarization 
for 31 protostellar sources buried in compact HII regions
and molecular clouds with the interstellar polarization 
in the vicinity of the sources as determined from field stars.
There was a strong tendency 
for the infrared and interstellar polarization vectors to be approximately parallel,
with sixty-five percent of the sample (29 sources) having position angles of polarization 
lying within 30$\arcdeg$ of the average local interstellar polarization direction.
NGC 2264 IRS 1 was one of the exceptions,
with the infrared polarization vector closer to the orientation of the galactic plane
than to the average interstellar polarization direction,
but is not close to either.
In the following, 
we compare near-infrared and optical polarization data
to reveal the relationship between the local and Galactic magnetic fields
in the IRS 2 region.

Interstellar polarizations in a (20$\arcdeg$ $\times$ 20$\arcdeg$) region
centered around IRS 2, were taken from the optical polarization data
in the Heiles catalogue (2000).
\footnote{An agglomeration of stellar polarization catalogs with results for 9286 stars}.
Figure 15 shows the relations 
between both polarization degrees and polarization angles against distances 
for the optical data.
It shows larger polarization degrees at larger distances, as expected,
as well as larger dispersion.
For the position angles there is a peak in the angles beyond the distance of 760 pc
in contrast to the sources in front of that.
Figure 16 shows histograms of the polarization position angles
for the local (our near-IR data) and Galactic magnetic field, and
Figure 17 shows the interstellar polarization vector map of the Galactic magnetic field
from optical polarization data.
The direction of the Galactic magnetic field runs 
roughly perpendicular to the direction of the local magnetic field.
In other words,
the local magnetic field is not aligned with the Galactic magnetic field.
Therefore, some other factors other than the Galactic magnetic field 
may have played a role in the formation and evolution 
of the cluster core in this region.
It is interesting to note that 
the IRAS 12 S1 core may be rotating (Wolf-Chase \& Walker 1995)
and that the suggested axis of rotation in NGC 2264 coincides with the axis of the Galaxy
(Dyck \& Lonsdale 1979).

\begin{figure}[!t]
\epsscale{1.0}
\plotone{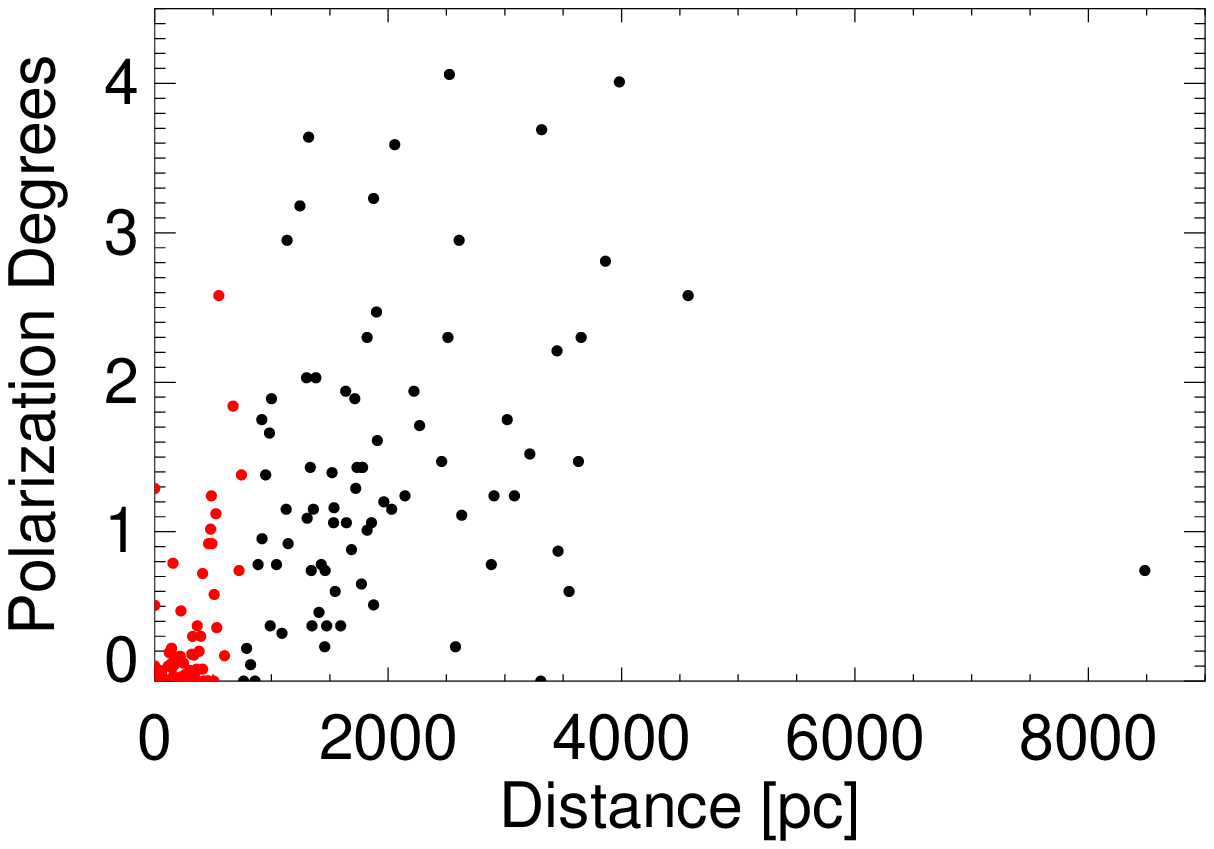}
\plotone{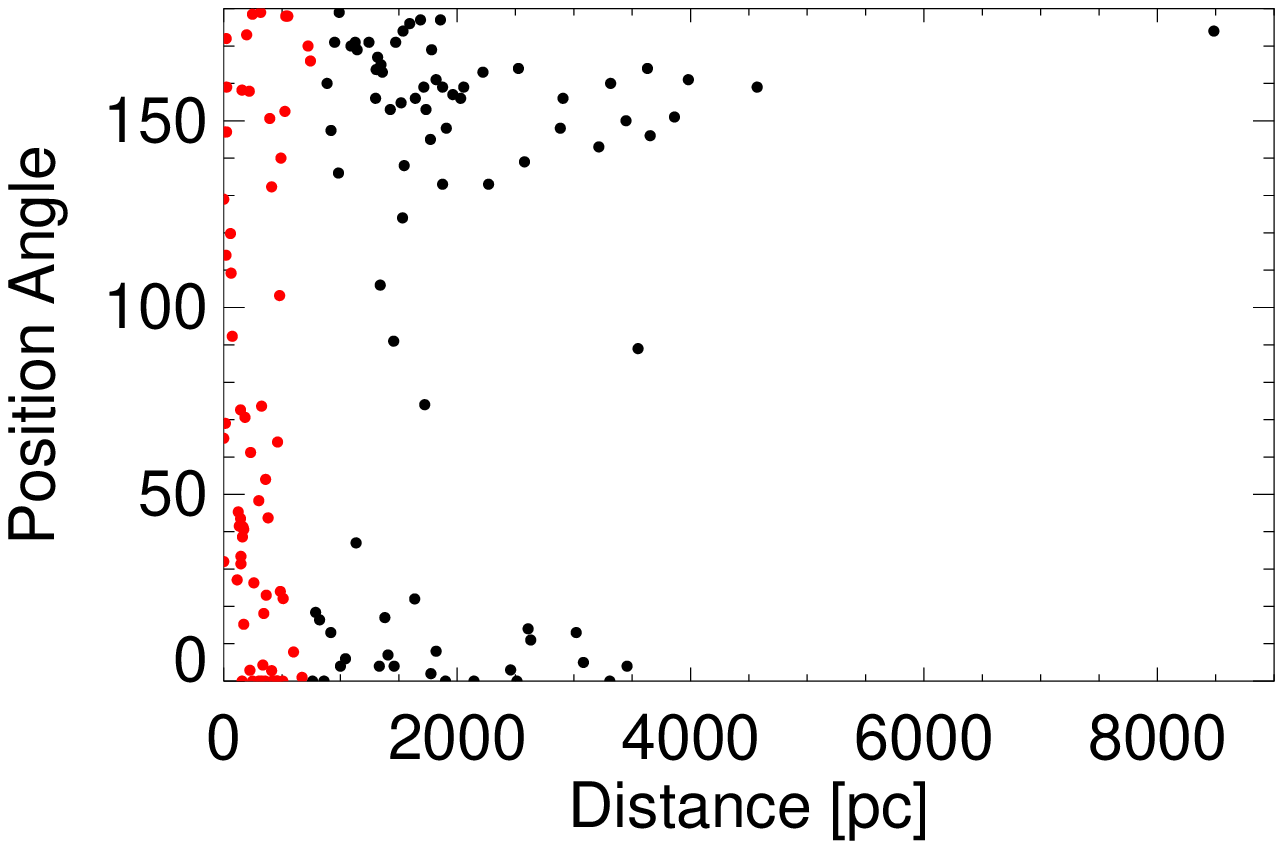}
\vspace{-0.5\baselineskip}
\vspace{-0.5\baselineskip}
\caption{
Polarization degrees (top panel) and
polarization position angles (bottom panel) against the distance
from the Heiles catalogue (2000).
Red circles: sources on below a distance of 760 pc.
Black circles: sources beyond a distance of 760 pc.}
\end{figure}

\begin{figure}[!t]
\epsscale{1.0}
\plotone{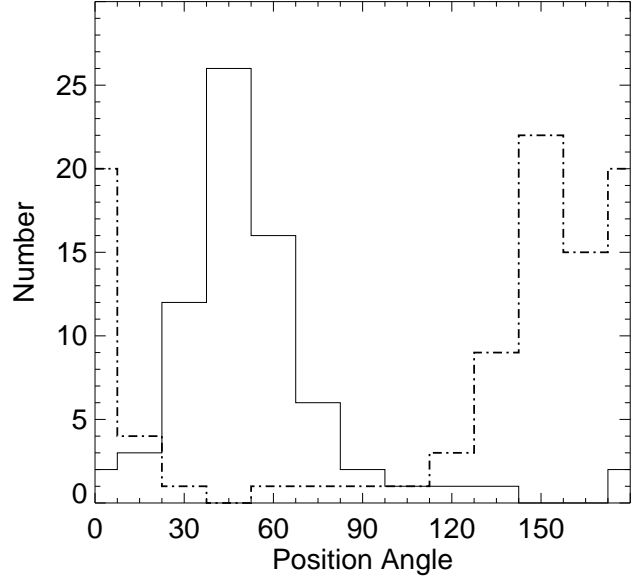}
\vspace{-0.5\baselineskip}
\vspace{-0.5\baselineskip}
\caption{
Histograms of polarization position angles.
Dot-dashed line:
polarization position angles of the sources from the Heiles catalogue (2000).
Solid line:
polarization position angles for the group Ar sources
in the perimeter of the cluster core for the $H$ band.}
\end{figure}

\subsubsubsection{4.2.3.2 Comparisons with Theory of Cluster Formation}

Very recently, Nakamura \& Li (2010) performed numerical simulations 
of clustered star formation in parsec-scale dense clumps
with different magnetic field strength 
and discussed the role of magnetic fields in cluster formation.
The initial cloud was a centrally condensed spherical clump
with a simulation box length of 2 pc.
The central density was 5.0 $\times$ 10$^{-20}$ g cm$^{-3}$ (1.5 $\times$ 10$^4$ cm$^{-3}$),
yielding a total clump mass of 884 M$_{\sun}$.
These physical quantities appear to be in a reasonable agreement
with those of the observed IRS 2 field.
They also assumed an isothermal equation of state with a sound speed of 0.23 km s$^{-1}$
for a mean molecular weight of 2.33 and a gas temperature of 15 K.
In addition, 
the magnetic field strength was specified by the plasma beta $\beta$,
the ratio of thermal pressure to magnetic pressure at the clump center,
through $B_0 = 25.8 \beta^{-1/2}$ $\mu$G.
In their numerical models it was found that
the filamentary clumps are almost perpendicular to the global magnetic fields 
when the cluster forming clumps are created out of a strongly-magnetized cloud.
In contrast, when the magnetic field associated with the parent molecular clouds
is dynamically weak,
the random component parallel to the filamentary clumps tends to be stronger 
preferentially in the dense parts 
when the turbulent flows play a role in the formation of the dense clumps.
The observed, spatially ordered magnetic field seems consistent with
their strongly-magnetized model 
for which the cloud dynamics is regulated by the strong magnetic field.
Therefore, our comparison between local and global magnetic fields
may indicate that the cluster forming clumps of the IRS 2 region 
including the protostars in the distribution corresponding to the Jeans length
may be created out of such a cloud,
in the presence of strong magnetic fields.
However, even in a magnetically subcritical cloud,
partly distorted field components exist.
This could explain the difference in the polarization position angles 
between the cluster core and the perimeter.

\begin{figure*}
\epsscale{1.9}
\plotone{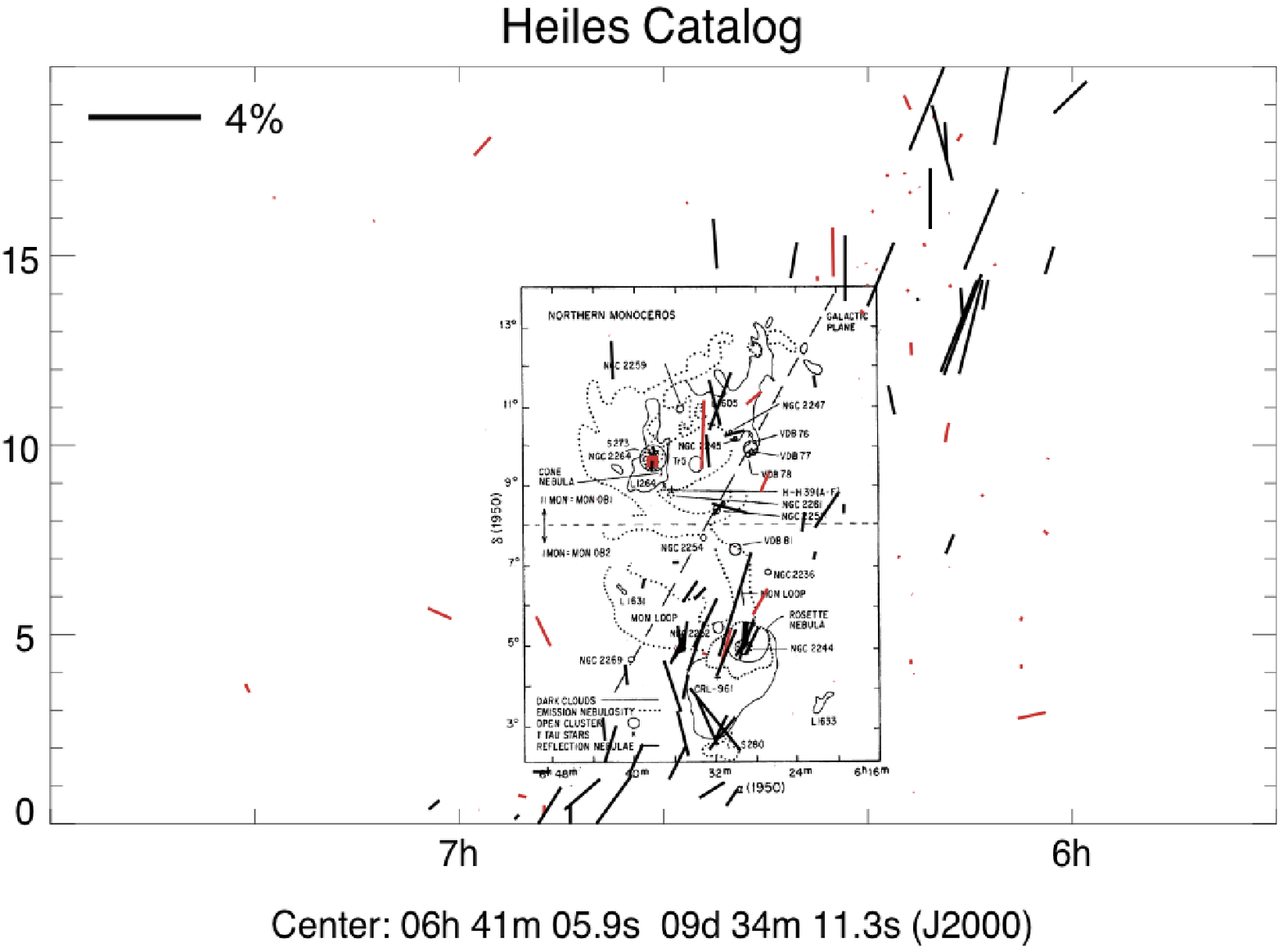}
\vspace{-0.5\baselineskip}
\vspace{-0.5\baselineskip}
\caption{
Comparison of the polarzation vector map from the Heiles catalogue (2000)
with the map of the Northern Monoceros region
including NGC 2264 (P\'{e}rez 1991).
The length of the vectors is proportional to the degree of polarization.
Shown in the upper left corner is a 4\% vector.
The red box is associated with the IRS 2 region.
Red lines: vectors for sources below a distance of 760 pc.
Black lines: vectors for sources beyond a distance of 760 pc.}
\end{figure*}

\subsubsection{Protostellar Outflows and Magnetic Fields}

Most cloud cores are magnetically supercritical
(Nakano 1998, 1999; Crutcher 1999; Shu et al. 1999),
and outflows are closely related to the magnetic field in the cloud
as well as the accretion process, angular momentum transportation.

Molecular outflows in the Monoceros OB1 cloud were reported
previously by Margulis \& Lada (1988).
The direction ($\sim$60$\arcdeg$) of the CO molecular outflow 
measured by connecting the peaks of the red- and blue-shifted components
is well aligned
with that of our observed local magnetic field in the IRS 2 region ($\sim$60$\arcdeg$).
Based on Wolf-Chase et al. (2003),
if IRAS 12 S1, a multiple source, is the main contributor to this outflow,
then the outflow momentum flux is roughly an order of magnitude 
greater than expected for a Class I object of comparable bolometric luminosity.
It means that IRAS 12 S1 consists of young Class 0 objects (Bontemps et al. 1996).

Outflow formation has been simulated by many researchers, 
as has protostellar collapse
(e.g., Tomisaka 1998, 2002; Machida et al. 2005a, 2005b, 2006;
Matsumoto \& Tomisaka 2004; Matsumoto et al. 2006; Ziegler 2005;
Banerjee \& Pudritz 2006; Fromang et al. 2006).
According to these simulations, outflow axes tend to be aligned
parallel to the local magnetic fields 
and perpendicular to rotating disks,
when the magnetic fields associated with parent cloud cores are not weak.
This implies that
the magnetic field strength in the IRS 2 region may be strong enough to align the outflows
in the direction of the local-scale magnetic field.

In addition, 
the NGC 2264 cluster contains stars at different stages of formation,
and the star formation process appears to have continued for a few Myrs
(Dahm \& Simon 2005),
corresponding to more than 10 free-fall times for a gas clump
with a few 10$^4$ cm$^{-3}$.
In other words,
the star formation in NGC 2264 is likely to be relatively slow.
Such slow star formation could be explained by protostellar outflow feedback
that can impede the global gravitational collapse 
by regenerating supersonic turbulence.
Recent numerical simulations have shown that
the star formation rate is significantly reduced 
to an observed low level particularly in the presence of moderately-strong magnetic field.
(e.g., Li \& Nakamura 2006; Nakamura \& Li 2007, 2010; Wang et al. 2010).

\subsubsection{Comparison with Submillimeter Polarimetry}

\begin{figure*}
\epsscale{1.9}
\plotone{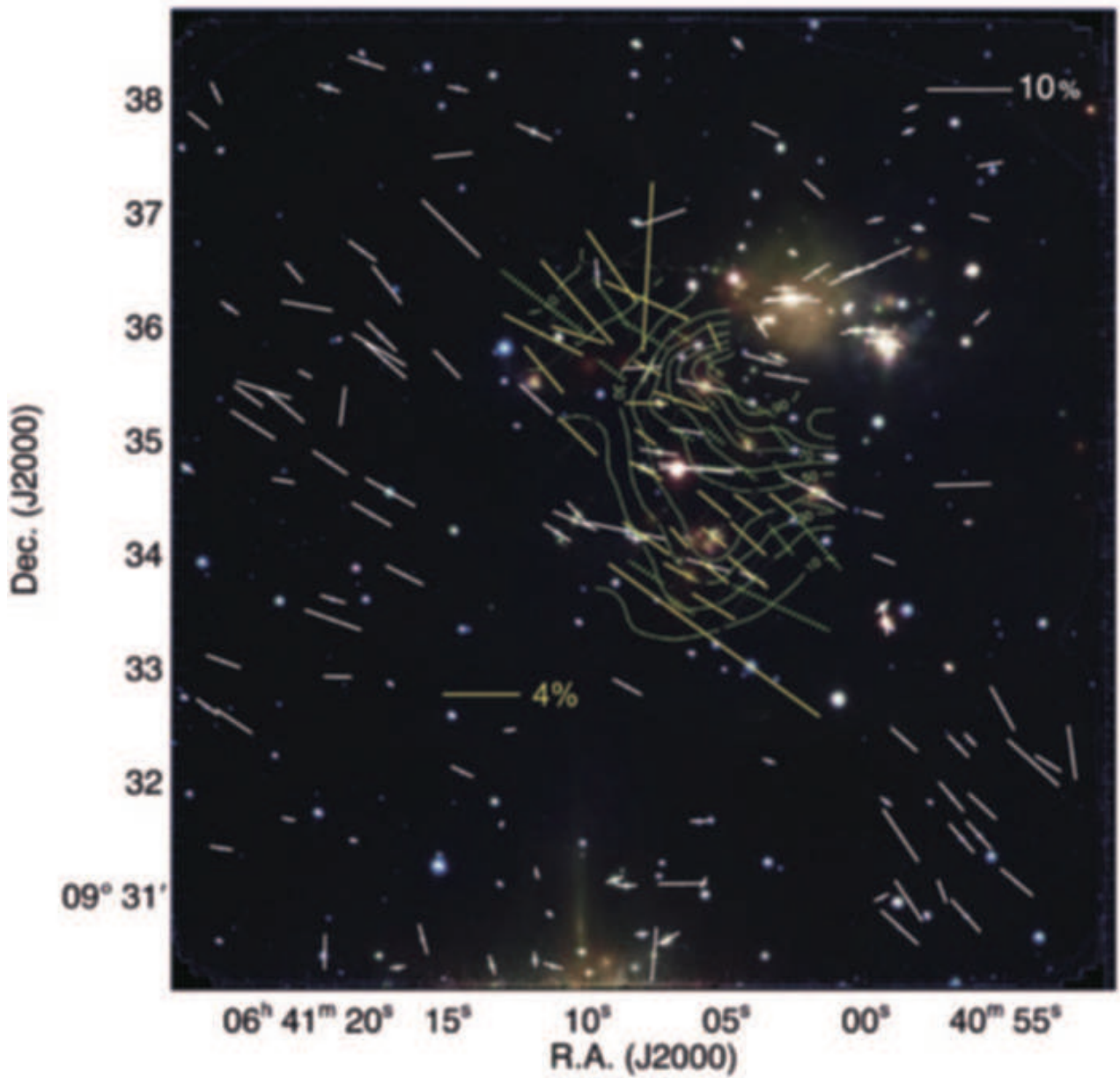}
\vspace{-0.5\baselineskip}
\vspace{-0.5\baselineskip}
\caption{
Submillimeter polarization vector map (Dotson et al. 2010)
superposed on $H$ polarization vector map.
Note that the submillimeter polarization vectors were rotated by 90$\arcdeg$ to show the inferred magnetic field direction and directly to compare our near-IR polarization vectors.
White vectors: $H$ polarization vectors.
Yellow vectors:
submillimeter polarization vectors (P/$\sigma$$_{p}$ $\ge$ 3).
Green dotted vectors:
submillimeter polarization vectors (2 $\le$ P/$\sigma$$_{p}$ $<$ 3).
Green contours: contours at 10\%, 20\%, ..., 90\% of the peak intensity of 62 Jy beam$^{-1}$.}
\end{figure*}

An interesting issue is 
how useful near-IR polarimetry is in tracing the magnetic field structure 
of dense clouds. 
Goodman et al. (1995) suggested that 
the polarizing power of dust grains may drop in the dense interior of some dark clouds 
and that near-IR polarization maps of background sources may be unreliable. 
However, the relevant physics is surprisingly complex (Lazarian 2007) 
and there are various pieces of observational evidence and theoretical explanations 
for aligned grains in dense cloud cores 
(Ward-Thompson et al. 2000; Cho \& Lazarian 2005; Hough et al. 2008). 
To compare the magnetic field direction
derived from dichroic polarization at near-IR wavelengths 
with that from thermal emission polarization at submillimeter wavelengths, 
it may be a worthwhile experiment to confirm the potential inefficiency 
of grain alignment in a dense region, 
even though previous observations have already shown agreement 
in the magnetic field structures at various wavelengths 
such as near-, far-IR, or submillimeter wavelengths 
(Tamura et al. 1996, 2007; Houde et al. 2004; Kandori et al. 2007).
Figure 18 shows a comparison with submillimeter polarization vector map 
(Dotson et al. 2010), 
and Figure 14 shows histograms of polarization position angles at $H$
and at a wavelength of 350 $\mu$m (Dotson et al. 2010). 
Interestingly, 
the distribution of the polarization position angles at 350 $\mu$m 
is in better agreement with that of the $H$ band 
in the perimeter of the cluster core than in the cluster core.
The mean angle of the polarization position angles at  350 $\mu$m is 60$\arcdeg$.

The general trend derived from the above two wavelengths 
represents a good agreement with each other, 
although their spatial scale is somewhat different. 
Therefore, our results indicate that 
both near-IR dichroic polarization and submillimeter emission polarization 
may trace the magnetic field structures associated with the IRS 2 region.

\subsubsection{Wavelength Dependence of Interstellar Polarization}

\begin{figure}[!t]
\epsscale{1.0}
\plotone{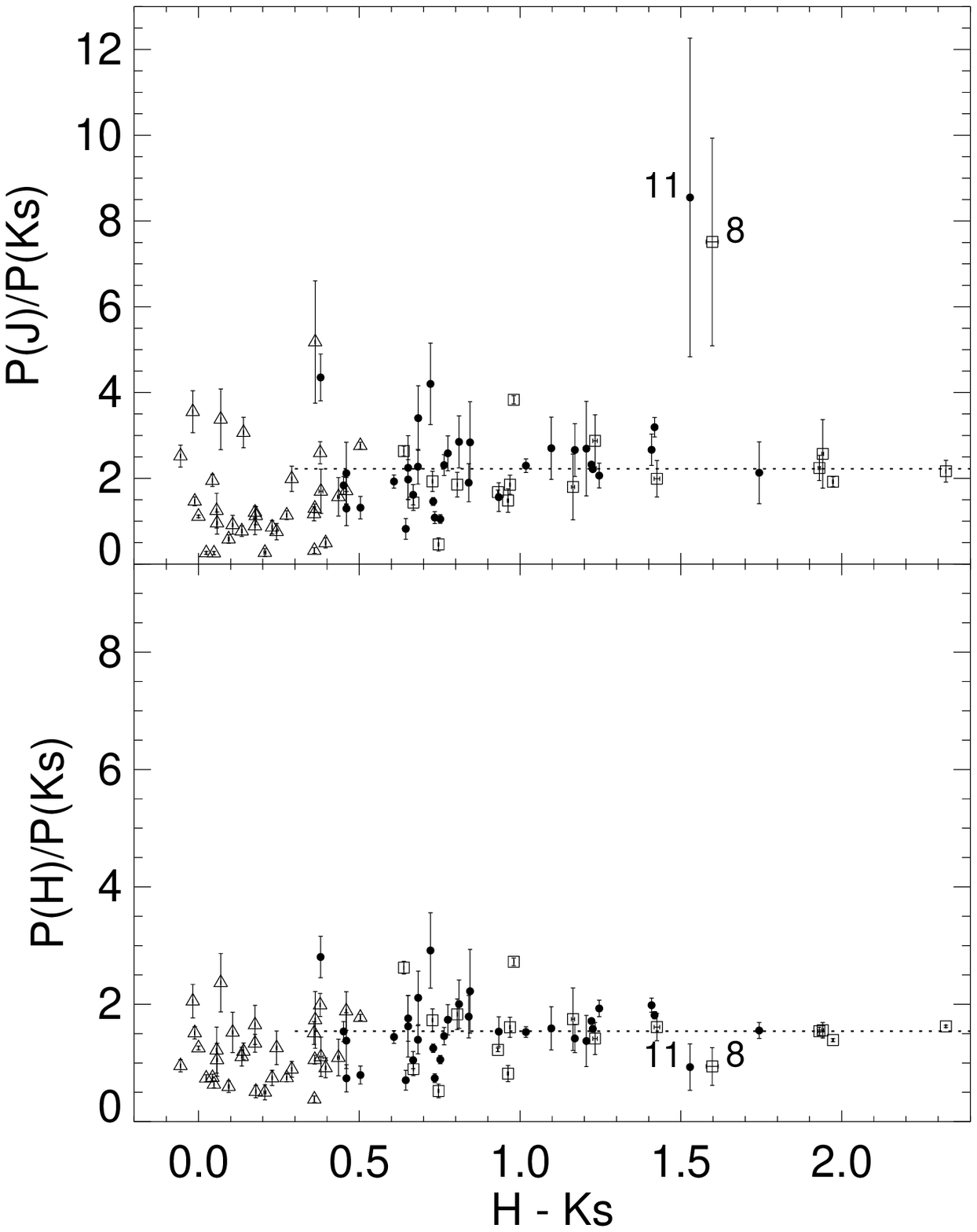}
\caption{
Ratio of $J$ to $K_s$ polarizations (top panel) 
and the ratio of $H$ to $K_s$ polarizations (bottom panel) 
against the $H - K_s$ color for the sources 
detected in the $J$, $H$, $K_s$ bands (Table 2). 
Triangles: sources in group A0.
Filled circles: sources in group Ar. 
Squares: sources in group B.
Dotted lines: weighted average for group Ar.}
\end{figure}

Though the near-IR polarizations of selected point-like sources are most likely dominated by dichroic extinction 
with no nebulosity, we cannot rule out the possibility of the presence of unresolved reflection nebulae
in some of the sources.
To discriminate between the contributions from the dichroic extinction and from the scattering,
the wavelength dependence of polarization can be measured by calculating the ratio of polarization degrees.
In the IR wavelength range,
the polarization by dichroic extinction decreases with wavelength, 
while the polarization by scattering is not a strong function of wavelength
(Whittet et al. 1992; Casali 1995).
Figure 19 shows the $P_J/P_{K_s}$ and $P_H/P_{K_s}$ ratios for the sources in the IRS 2 region.
It is very clear that group A0 and Ar show very different behavior.
Within group Ar, the ratios are reasonably constant, 
and the weighted average values are $P_J/P_{K_s}$ = 2.22 and $P_H/P_{K_s}$ = 1.54 
with standard deviations of 0.80 and 0.50, respectively.
These values are consistent with the empirical relation $P \propto \lambda ^{-\beta}$ = 1.6--2.0
(Whittet 1992).
Therefore, the polarization of group Ar sources can be very well explained by dichroic extinction.
Source 11 shows an unusually high ratio in Figure 19 (top panel).
The $J$ polarization degree of source 11 is too high with large errors,
while the $P_H/P_{K_s}$ ratio is near 1 as expected.
Since source 11 is located around boundary, 
it may be possible that it was caused by a bad pixel in the $J$ band.

In contrast, the $P_J/P_{K_s}$ and $P_H/P_{K_s}$ ratios of group A0 sources are near unity,
which is significantly different from those of group Ar sources.
Therefore, the polarization mechanism for low extinction sources may be dominated by the circumstellar scattering
(see more discussions in Section 5.4 of Casali 1995).

The $P_J/P_{K_s}$ and $P_H/P_{K_s}$ ratios of group B are similar to that of group Ar,
and Source 8 also shows an unusually high ratio in Figure 19 (top panel).
The $J$ polarization degree is too high with large errors,
while the $P_H/P_{K_s}$ ratio is near 1 as expected.
It is the same as source 11.

\section{SUMMARY}
We conducted deep and wide-field $J$-$H$-$K_s$ imaging polarimetry
In the direction of a 8$\arcmin$ $\times$ 8$\arcmin$ region around NGC 2264 IRS 2
in the Monoceros OB1 cloud.
The main results in this study are summarized as follows.

\begin{enumerate}

\item
Our polarization data revealed three IRNCs associated with IRS 2, 
and several local IRNe. In addition, the illuminating sources of the IRNe 
were identified with near- and mid-IR sources.

\item
Aperture photometry of point-like sources, with 314 detected in all three bands,
was used to classify sources using a color--color diagram.

\item
Aperture polarimetry of the point-like sources 
allowed positive detection of 241 sources
in at least one of the three bands. 
Most of the near-IR polarizations of the point-like sources
can be explained by dichroic polarization.

\item
Sources in group Ar are expected to be either background stars or PMS stars
with little intrinsic polarization.
The histogram of polarization position angles of group Ar sources in the cluster core
has a well-defined peak at $\sim$81$\arcdeg$,
which we interpret as the projected direction, on the sky, of the magnetic fields in the IRS 2 region.
From the 29$\arcdeg$ standard deviation of the polarization position angles,
the strength of the magnetic field projected on the plane of the sky
is roughly estimated at $\sim$100 $\mu$G using the Chandrasekhar--Fermi method.

\item
By comparing the observed magnetic field lines 
and those derived from recent numerical simulations,
we suggest that the clump including the IRS 2 region is likely to have formed
under the influence of a strong magnetic field.
The local magnetic field runs roughly perpendicular to the Galactic magnetic field.
The significant difference between the local and Galactic magnetic field directions
may imply that the local magnetic field is strong enough to control the cloud dynamics.
In fact, the direction of the powerful molecular outflow from IRAS 12 S1
is along the local magnetic field,
instead of the Galactic magnetic field.

\item
The magnetic field direction inferred from our observations appears 
to be consistent with that inferred from 350 $\mu$m thermal continuum emission polarimetry 
centered around IRAS 12 S1.
The 350 $\mu$m polarization position angles were especially consistent with
the group Ar sources in the outer parts of the cluster core.

\item
For the group Ar sources, 
the wavelength dependence of polarization is consistent with the dichroic extinction.
Sources in group A0 have a small amount of extinction,
and their polarization seems to be caused by the circumstellar scattering.
Sources in group B show similar behavior to the group Ar sources in our data.

\end{enumerate}

\acknowledgments
This work was supported by the Korean Scholarship Foundation. 
M. T. has been supported by the MEXT, Grants-in-Aid 19204018 and 22000005. 
This work is based on observations made at the South African Astronomical Observatory 
(Department of Astronomy, Nagoya University). 
This work is based in part on observations made with the Spitzer Space Telescope,
which is operated by the Jet Propulsion Laboratory, 
California Institute of Technology under a contract with NASA. 
This publication makes use of data products from the Two Micron All Sky Survey, 
which is a joint project of the University of Massachusetts and 
the Infrared Processing and Analysis Center/California Institute of Technology, 
funded by the National Aeronautics and Space Administration and the National Science Foundation.
This research has made use of the SIMBAD database, operated at CDS, Strasbourg, France,
as well as IRAF and the IDL Astronomy Library.

\clearpage

\begin{deluxetable}{lccccccll}
\rotate
\tabletypesize{\tiny} 
\tablewidth{0pt}
\tablecaption{Photometry of Point-like Sources in the Spokes Cluster}
\tablehead{
\colhead{Source} &\multicolumn{2}{c}{Position} & \colhead{2MASS} & \colhead{$J$} & \colhead{$H$} & \colhead{$K_s$}
& \colhead{Group\tablenotemark{a}} &\colhead{note} \\
\cline{2-3}
& \colhead {$\alpha_{\rm J2000.0}$} & \colhead {$\delta_{\rm J2000.0}$}
& \colhead {} & \colhead{(mag)} & \colhead{(mag)} & \colhead{(mag)} & \colhead {} & }
\startdata
           1 & 6 41 16.79 & 9 30 22.5 & J06411679+0930225 & 15.655 & 13.819 & 12.850 & Ar  & FMS 2-1404,                                         LBM 6274   \\
           2 & 6 41 11.21 & 9 30 23.3 & J06411121+0930235 & 14.651 & 14.068 & 13.823 & A0  & FMS 2-1308, FMS 397,                                LBM 6112,           RMS 3667  \\
           3 & 6 41 13.40 & 9 30 23.8 & J06411340+0930237 & 14.554 & 13.362 & 12.606 & B   & FMS 2-1355,                                         LBM 6182,           RMS 3727  \\
           4 & 6 41 11.85 & 9 30 26.7 & J06411185+0930267 & 14.354 & 13.332 & 12.807 & Ar  & FMS 2-1321,                                         LBM 6128,           RMS 3680 \\
           5 & 6 41 19.45 & 9 30 28.7 & J06411945+0930286 & 12.195 & 11.779 & 11.728 & A0  & FMS 2-1464, FMS 431, [FMS2006] 374,                 LBM 6390, LBM 6391, RMS 3862, PSB 173 \\
             &            &           &                   &        &        &        &     &  SBL 352, VAS 154, [NGC 2264 454, GEN +2.22640454   \\
           6 & 6 41 17.44 & 9 30 28.7 & J06411744+0930287 & 14.378 & 12.753 & 11.847 & Ar  & FMS 2-1422,          [FMS2006] 357,                 LBM 6308,           RMS 3816  \\
           7 & 6 41 03.61 & 9 30 29.0 & J06410360+0930290 & 12.780 & 12.049 & 11.699 & A0  & FMS 2-1022, FMS 334, [FMS2006] 207,                 LBM 5810,           RMS 3447, PSB 1467, DS 302 \\
\enddata
\end{deluxetable}
\begin{deluxetable}{lrrrrrr}
\tabletypesize{\scriptsize}
\tablewidth{0pt}
\tablecaption{Polarimetry of Point-like Sources in the Spokes Cluster}
\tablehead{
\colhead{Source} & \colhead{$P_J$} & \colhead{$P_H$} & \colhead{$P_{K_s}$}
& \colhead{$\theta_J$} & \colhead{$\theta_H$} & \colhead{$\theta_{K_s}$} \\
& \colhead{(\%)} & \colhead{(\%)} & \colhead{(\%)}
& \colhead{($\arcdeg$)} & \colhead{($\arcdeg$)} & \colhead{($\arcdeg$)} }%
\startdata
1       &               \mcoc{$<$               6.9     }       &       1.7     $\pm$   0.5                     &               \mcoc{$<$               4.8     }       &               \mcoc{\ldots}           &       98.4    $\pm$   7.5     &               \mcoc{\ldots}           \\
2       &               \mcoc{$<$               2.8     }       &       2.3     $\pm$   0.5                     &               \mcoc{$<$               10.1    }       &               \mcoc{\ldots}           &       73.6    $\pm$   6.6     &               \mcoc{\ldots}           \\
3       &               \mcoc{$<$               2.4     }       &       2.3     $\pm$   0.3                     &               \mcoc{$<$               3.1     }       &               \mcoc{\ldots}           &       17.3    $\pm$   3.5     &               \mcoc{\ldots}           \\
4       &       2.8     $\pm$   0.6                     &       1.8     $\pm$   0.2                     &               \mcoc{$<$               3.1     }       &       20.8    $\pm$   5.8     &       5.3     $\pm$   3.8     &               \mcoc{\ldots}           \\
5       &               \mcoc{$<$               0.22    }       &       0.25    $\pm$   0.06                    &               \mcoc{$<$               1.0     }       &               \mcoc{\ldots}           &       128.4   $\pm$   6.5     &               \mcoc{\ldots}           \\
6       &               \mcoc{$<$               1.4     }       &       0.54    $\pm$   0.12                    &               \mcoc{$<$               1.13    }       &               \mcoc{\ldots}           &       144.8   $\pm$   6.2     &               \mcoc{\ldots}           \\
7       &       0.61    $\pm$   0.12                    &       0.23    $\pm$   0.07                    &               \mcoc{$<$               0.9     }       &       82.3    $\pm$   5.3     &       178.8   $\pm$   7.8     &               \mcoc{\ldots}           \\
\enddata
\tablecomments{For sources with $P/\delta P < 3$, the 3$\delta P$ upper limits are listed.}
\end{deluxetable}


\begin{references}

\reference{} Adams, M. T., Strom, K. M., \& Strom, S. E. 1979, \apj, 230, L183
\reference{} Allen, D. A. 1972, \apj, 172, L55
\reference{} Alves, F. O., Franco, G. A. P., \& Girart, J. M. 2008,
             \aap, 486, L13
\reference{} Andersson, B.-G., \& Potter, S. B. 2005, \mnras, 356, 1088
\reference{} Aspin, C., \& Reipurth, B. 2003, \aj, 126, 2936
\reference{} Banerjee, R., \&  Pudritz, R. E. 2006, \apj, 641, 949
\reference{} Bastien, P., \& Menard, F. 1988, \apj, 326, 334
\reference{} Bastien, P., \& Menard, F. 1990, \apj, 364, 232 
\reference{} Baxter, E. J., Covey, K. R., Muench, A. A., F\H{u}r\'{e}sz, G., Rebull, L., Szentgyorgyi, A. H.
             2009, \aj, 138, 963
\reference{} Bourke, T. L., Myers, P. C., Robinson, G., \& Hyland, A. R. 2001, \apj, 554, 916
\reference{} Breger, M., \& Hardorp, J. 1973, \apj, 183, L77
\reference{} Casali, M. M. 1995, MNRAS, 277, 1385
\reference{} Caselli, P., Benson, P. J., Myers, P. C., \& Tafalla, M.
              2002, \apj, 572, 238
\reference{} Castelaz, M. W. \& Grasdalen, G. 1988, \apj, 335, 150
\reference{} Chandrasekhar, S., \& Fermi, E. 1953, \apj, 118, 113
\reference{} Cho, J., \& Lazarian, A. 2005, \apj, 631, 361
\reference{} Cohen, M., Harvey, P. M., Wilking, B. A., \& Schwartz, R. D. 1984, \apj, 278, 671
\reference{} Crutcher, R. M. 1999, \apj, 520, 706
\reference{} Dahm, S. E., \& Simon, T. 2005, \aj, 129, 829
\reference{} Dahm, S. E. 2008, in Handbook of Star Forming Regions Vol. I
\reference{} Dotson, J. L., Vaillancourt, J. E., Kirby, L., Dowell, C. D., Hildebrand, R. H., 
	      \& Davidson, J. A. 2010, \apj, 186, 406
\reference{} Dyck, H. M., \& Lonsdale, C. J. 1979, \aj, 84,1339
\reference{} Forbrich, J., Tappe, A, Robitaille, T., Muench, A. A., Teixeira, P. S., 
	      Lada, E. A., Stolte, A., \& Lada, C. J. 2010, \apj, 716, 1453
\reference{} Fromang, S., Hennebelle, P., \& Teyssier, R. 2006, \aap, 457, 371
\reference{} Goodman, A. A., Benson, P. J., Fuller, G. A., \& Myers, P. C.	
	      1993, \apj, 406, 528
\reference{} Goodman, A. A., Jones, T. J., Lada, E. A., \& Myers, P. C. 
	      1995, \apj, 448, 748
\reference{} Hashimoto, J., Tamura, M., Kandori, R., Kusakabe, N., Nakajima, Y., Kurita, M., 
	      Nagata, T., Nagayama, T., Hough, J., \& Chrysostomou, A. 2008, \apj, 677, L39
\reference{} Heckert, P. A., \& Zeilik, M., II. 1981, \aj, 86, 1076
\reference{} Heckert, P. A., \& Zeilik, M. 1984, \aj, 89, 1379
\reference{} Heiles, C. 2000, \aj, 119, 923
\reference{} Herbig, G. H. 1954, \apj, 119, 483
\reference{} Herbig, G. H. 1974, \apj, Lick Obs. Bull., No. 658
\reference{} Hodapp, K.-W. 1994, \apj, 94, 615
\reference{} Holland, W. S., Greaves, J. S., Ward-Thompson, D., \& Andre, P. 
	      1996, \aap, 309, 267
\reference{} Houde, M., Dowell, C. D., Hildebrand, R. H., Dotson, J. L., Vaillancourt, J. E.,
              Phillips, T. G., Peng, R., \& Bastien, P.
              2004, \apj, 604, 717
\reference{} Hough, J. H., Aitken, D. K., Whittet, D. C. B., Adamson, A. J., \& Chrysostomou, A.
              2008, \mnras, 387, 797
\reference{} Jones, T. J. 1989, \apj, 346, 728
\reference{} Jones, T. J., \& Amini, H. 2003, \aj, 125, 1418
\reference{} Kandori, R., et al. 2006, Proc. SPIE, 6269, 159
\reference{} Kandori, R., Tamura, M., Kusakabe, N., Nakajima, Y., Nagayama, T., 
	      Nagashima, C., Hashimoto, J., Ishihara, A., Nagata, T., Hough, J. H.
	      2007, \pasj, 59, 487
\reference{} Kobayashi, Y., Kawara, K., Maihara, T., Okuda, H., Sato, S., \& Noguchi, K. 1978, \pasj, 30, 377
\reference{} Kr\"{u}gel, E., Guesten, R., Schulz, A., \& Thum, C. 1987, \aap, 185, 283
\reference{} Kwon, J., Choi, M., Pak, S., Kandori, R., Tamura, M., Nagata, T., \& Sato, S.
	      2010, \apj, 708, 758
\reference{} Lada, C. J., \& Adams, F. C. 1992, \apj, 393, 278
\reference{} Larson, R. B. 1969, \mnras, 145, 271
\reference{} Lazarian, A. 2007, J. Quant. Spectrosc. Radiat. Transfer, 106, 225
\reference{} Li, Z.-Y., \& Nakamura, F. 2006, \apj, 640, 187
\reference{} Machida, M. N., Matsumoto, T., Tomisaka, K., \& Hanawa, T. 
	      2005a, \mnras, 362, 369
\reference{} Machida, M. N., Matsumoto, T., Hanawa, T., \& Tomisaka, K.
	      2005b, \mnras, 362, 382
\reference{} Machida, M. N., Inutsuka, S.-I., Matsumoto, T. 2006, \apj, 647, L151
\reference{} Margulis, M., Lada, C. J., \& Snell, R. L. 1988, \apj, 333, 316
\reference{} Margulis, M., Lada, C. J., \& Young, E. T. 1989, \apj, 345, 906
\reference{} Matsumoto, T., \& Tomisaka, K. 2004, \apj, 616, 266
\reference{} Matsumoto, T., Nakazato, T., \& Tomisaka, K. 2006, \apj, 637, L105
\reference{} Nagayama, T., Nagashima, C., Nakajima, Y., Nagata, T., Sato, S., Nakaya, H., Yamamuro, T., Sugitani, K., \& Tamura, M. 2003, Proc. SPIE, 4841, 459
\reference{} Nakamura, F., \& Li, Z.-Y. 2007, \apj, 662, 395
\reference{} Nakamura, F., \& Li, Z.-Y. 2010, in Proceedings IAU Symposium No. 270
\reference{} Nakano, T. 1998, \apj, 494, 587
\reference{} Nakano, T. 1999, in Proceedings of Star Formation 1999
\reference{} Nakano, T., Nishi, R., \& Umebayashi, T. 2002, \apj, 573, 199
\reference{} Ostriker, E. C., Stone, J. M., \& Gammie, C. F. 2001,
	      \apj, 546, 980
\reference{} Peretto, N., Andr\'{e}, Ph., \& Belloche, A. 2006, \aap, 445, 979
\reference{} P\'{e}rez, M. R., The, P. S., \& Westerlund, B. E. 1987, \pasp, 99, 1050
\reference{} Poidevin, F., \& Bastien, P. 2006, ApJ, 650, 945
\reference{} Reipurth, B., Yu, K. C., Moriarty-Schieven, G., Bally, J., Aspin, C., 
	      \& Heathcote, S. 2004, \aj, 127, 1969
\reference{} Sargent, A. I., Van Duinen, R. J., Nordh, H. L., Fridlund, C. V. M.,
	      Aalders, J. W., \& Beintema, D. 1984, \aap, 135, 377
\reference{} Schreyer, K., Stecklum, B., Linz, H., \& Henning, Th. 2003, \apj, 599, 335
\reference{} Schwartz, P. R., Thronson, H. A., Jr., Odenwald, S. F., Glaccum, W., Loewenstein, R. F., Wolf, G. 1985, \apj, 292, 231
\reference{} Shu, F. H., Allen, A., Shang, H., Ostriker, E. C., Li, Z.-Y. 1999,
	      The Origin of Stars and Planetary Systems. 
	      Edited by Charles J. Lada and Nikolaos D. Kylafis.
\reference{} Stetson, P. B. 1987, \pasp, 99, 191
\reference{} Strom, K. M., Strom, S. E., Wolff, S. C., Morgan, J. \& Wenz, M. 1986, \apjs, 62, 39
\reference{} Sugitani, K., Nakamura, F., Tamura, M., Watanabe, M., Kandori, R., 
	      Nishiyama, S., Kusakabe, N., Hashimoto, J., Nagata, T., \& Sato, S.
	      2010, \apj, 716, 299
\reference{} Sung, H., Bessell, M. S., \& Lee, S.-W. 1997, \aj, 114, 2644 
\reference{} Sung, H., \& Bessell, M. S. 2010, \aj, 140, 2070
\reference{} Tamura, M., \& Sato, S. 1989, \aj, 98, 1368
\reference{} Tamura, M., Gatley, Ian, Joyce, R. R., Ueno, M., Suto, H. \& Sekiguchi, M. 1991, \apj, 378, 611
\reference{} Tamura, M., Hayashi, S., Itoh, Y., Hough, J. H., \& Chrysostomou, A.
	      1996, ASPC, 97, 372
\reference{} Tamura, M., et al. 2007, \pasj, 59, 467
\reference{} Teixeira, P. S., et al. 2006, \apj, 636, L45
\reference{} Teixeira, P. S., Zapata, L. A., \& Lada, C. J. 2007, \apj, 667, L179
\reference{} Tomisaka, K. 1998, \apj, 502, L163
\reference{} Tomisaka, K. 2002, \apj, 575, 306
\reference{} Vrba, F. J. 1977, \aj, 82, 198
\reference{} Vrba, F. J., Luginbuhl, C. B., Strom, S. E., Strom, K. M., \& Heyer, M. H. 1986, \aj, 92, 633
\reference{} Vrba, F. J., Strom, S. E., \& Strom, K. M. 1988, \aj, 96, 680
\reference{} Walker, M. F. 1954, \aj, 59, 333
\reference{} Walker, M. F. 1956, \apjs, 2, 365
\reference{} Walsh, J. R., Ogura, K., \& Reipurth, B. 1992, \mnras, 257, 110
\reference{} Wang, P., Li, Z.-Y., Abel, T. \& Nakamura, F. 2010, \apj, 709, 27
\reference{} Wardle, J. F. C., \& Kronberg, P. P. 1974, \apj, 194, 249
\reference{} Ward-Thompson, D., Kirk, J. M., Crutcher, R. M., Greaves, J. S.,
             Holland, W. S., \& Andr{\'e}, P. 2000, \apj, 537, L135
\reference{} Whittet, D. C. B. 1992, Dust in the Galactic Environment (Bristol: Institute of Physics Publishing)
\reference{} Whittet, D. C. B., Martin, P. G., Hough, J. H., Rouse, M. F., Bailey, J. A., \& Axon, D. J.
             1992, \apj, 386, 562
\reference{} Williams, J. P., \& Garland, Catherine A. 2002, \apj, 568, 259
\reference{} Wolf-Chase, G. A., \& Walker, C. K. 1995, \apj, 447, 244
\reference{} Wolf-Chase, G., Moriarty-Schieven, G., Fich, M., \& Barsony, M. 
	      2003, \mnras, 344, 809
\reference{} Young, E. T., Teixeira, P. S., Lada, C. J., Muzerolle, J., Persson, S. E., 
              Murphy, D. C., Siegler, N., Merengo, M., Krause, O., 
              \& Mainzer, A. K. 2006, \apj, 642, 972
\reference{} Ziegler, U. 2005, \aap, 435, 385


\end{references}
\end{document}